\documentclass[onecolumn,preprint,superscriptaddress]{revtex4-1}
\usepackage{bibentry,natbib}
\usepackage{latexsym,bm}
\usepackage{graphicx}
\usepackage{rotating}
\usepackage{amsmath}
\usepackage{rotating}
\usepackage{color}
\usepackage[colorlinks,citecolor=red]{hyperref}
\usepackage{subfigure}%??????
\usepackage{ulem}
\DeclareGraphicsExtensions{.jpg}

\begin{document}
\title{Three-body recombination near the d-wave resonance in ultracold $^{85}$Rb\,-$^{87}$Rb mixtures  }
\author{Cai-Yun Zhao}
\affiliation{State Key Laboratory of Magnetic Resonance and Atomic and Molecular Physics, Wuhan Institute of Physics and Mathematics, Chinese Academy of Sciences, Wuhan 430071, P. R. China}
\affiliation{ University of Chinese Academy of Sciences, 100049, Beijing, P. R. China}
\author{Hui-Li Han}
\email{huilihan@wipm.ac.cn}
\affiliation{State Key Laboratory of Magnetic Resonance and Atomic and Molecular Physics, Wuhan Institute of Physics and Mathematics, Chinese Academy of Sciences, Wuhan 430071, P. R. China}
\author{Ting-Yun Shi}
%\email{tyshi@wipm.ac.cn}
\affiliation{State Key Laboratory of Magnetic Resonance and Atomic and Molecular Physics, Wuhan Institute of Physics and Mathematics, Chinese Academy of Sciences, Wuhan 430071, P. R. China}
\date{\today}

\begin{abstract}
In this study, we investigate the three-body recombination (TBR) rates on both sides of the interspecies d-wave Feshbach resonance in the $^{85}$Rb\,-$^{87}$Rb-$^{87}$Rb system using the $R$-matrix propagation method in the hyperspherical coordinate frame. Our calculations are based on the Lennard-Jones model potential for the Rb-Rb interaction. Two different mechanisms of recombination rate enhancement for positive and negative $^{85}$Rb\,-$^{87}$Rb d-wave scattering lengths are analyzed. On the positive scattering length side, recombination rate enhancement occurs due to the existence of three-body shape resonance, while on the negative scattering length side, the coupling between the lowest entrance channel and the highest recombination channel is crucial to the appearance of the enhancement. In addition, our study shows that the intraspecies interaction has a significant role in determining the emergence of recombination rate enhancements. Compared to the case in which the three pairwise interactions are in d-wave resonance, when the $^{87}$Rb-$^{87}$Rb interaction is near the d-wave resonance, the values of the interspecies scattering length that produce the recombination enhancement shift. In particular, when the $^{87}$Rb-$^{87}$Rb interaction is away from the d-wave resonance, the enhancement disappears on the negative interspecies scattering length side.

\end{abstract}
\maketitle
\section{Introduction}
Three-body recombination occurs in ultracold atomic gases as a result of a three-body collision, in which the atoms gain kinetic energy due to the formation of a two-body bound state. This process is vitally important in a number of physical and chemical contexts and has been recognized as one of the most important scattering observables\;\cite{PhysRevLett.83.1566,PhysRevLett.83.1751,PhysRevLett.94.213201,BRAATEN2006259,efimov_phisics}. As an exothermic reaction, TBR is one of the main loss mechanisms in systems with ultracold atoms, limiting the density and lifetime of a Bose-Einstein condensate\;\cite{PhysRevLett.79.337,PhysRevLett.85.728,PhysRevLett.91.123201}. Additionally, the recombination process is employed as a way to form weakly bound diatoms in ultracold degenerate Fermi gases\;\cite{PhysRevLett.91.240401,PhysRevLett.91.240402,PhysRevLett.91.250401}.

Particles with resonant s-wave interactions will exhibit the Efimov effect, i.e., an infinite sequence of universal bound states characterized by discrete scale invariance\;\cite{Efimov1971,Efimov1973}. The universal properties of the Efimov effect have been investigated both theoretically\;\cite{jiawangPRL2012,yujunwangPRl2012,Sorensen2012,Wumengshan2014,Huang2014Li6,Wumengshan2016,
Hafner2017,Hanhuili2018,Wenz2009Li6,PhysRevA.100.052702}
and experimentally\;\cite{Kraemer2006Cs,Zaccanti2009Cs,Berninger2011Cs,Huang2014Cs,
Lompe2010Li6,Williams2009Li6,Huckans2009Li6,Ottenstein2008Li6,Pollack2009Li7,Gross2009Li7,Gross2010Li7,
Kunitski2015He,
Barontini2009rb87k41,Bloom2013rb87k40,HuMingGuang2014rb87k40,Wacker2016rb87k3941,Kato2017rb87k4041,Maier2015rb87Li7,Pires2014cslia-,
Tung2014cslia-,Ulmanis2016cslia+,Ulmanis_Kuhnle2016csli,Johansen2017csli,PhysRevLett.125.243401} in ultracold atomic gases. Due to the novel characteristic of the Efimov effect in few-body physics, whether the Efimov effect is possible for p-wave or higher partial-wave interaction is fundamentally important\;\cite{PhysRevLett.97.023201,PhysRevA.86.012711}. Nisida\;\cite{PhysRevA.86.012710} noted that Efimov states cannot be realized in physical situations for non-s-wave interactions. The studies of Efremov \textit{et al.}\;\cite{PhysRevLett.111.113201} indicated that the effective potential is attractive and decreases as the third power of the interatomic distance for a heavy-heavy-light system with a p-wave resonant interaction when employing the Born-Oppenheimer approximation method. With the same method, Zhu and Tan\;\cite{PhysRevA.87.063629} provided a more general discussion of the universal properties for atoms near higher partial-wave Feshbach resonances and found that the effective potential behaved as $1/\rho^{2L+1}$ when the distance $\rho$ between two heavy atoms was large. These two works also demonstrated that the Efimov effect does not occur in few-body systems interacting via higher partial-wave resonant interactions.

For d-wave dimers, Gao\;\cite{PhysRevA.62.050702} predicted that the bound states always appear at a universal value of the s-wave scattering length of $a_s\approx0.956\,r_{\scriptscriptstyle\textsl{vdW}}$. $r_{\scriptscriptstyle\textsl{vdW}}=(2\mu_{2b}C_6)^{1/4}/2$ is the van der Waals length, which characterizes the length scale between the two neutral atoms interacting via the van der Waals interaction $-C_6/r^6$ with two-body reduced mass $\mu_{2b}$. For a system of three identical bosons that interact via a d-wave resonant interaction, the study of Wang \textit{et al.}\;\cite{PhysRevA.86.062511} showed that a universal three-body state associated with the d-wave dimer is formed at $a_s\approx1.09\,r_{\scriptscriptstyle\textsl{vdW}}$. Calculations by Juan \textit{et al.}\;\cite{PhysRevA.99.012701} showed that the TBR rate monotonically increases through the unitary point and is nearly a constant on the quasi-bound side. Despite this progress, universal properties of heteronuclear systems interacting via resonant higher partial-wave interactions are less well understood\;\cite{PhysRevLett.120.023401}.

Experimentally, great efforts have been devoted to studying strongly interacting atomic Bose gases with s-wave resonances\;\cite{Kraemer2006Cs,Zaccanti2009Cs,Berninger2011Cs,Huang2014Cs,
Pollack2009Li7,Gross2009Li7,Gross2010Li7,Kunitski2015He,Barontini2009rb87k41,Bloom2013rb87k40,HuMingGuang2014rb87k40,Wacker2016rb87k3941,Kato2017rb87k4041,Maier2015rb87Li7,Pires2014cslia-,
Tung2014cslia-,Ulmanis2016cslia+,Ulmanis_Kuhnle2016csli,Johansen2017csli,PhysRevLett.110.163202,PhysRevLett.123.233402,PhysRevLett.125.243401,PhysRevLett.111.125303,PhysRevX.6.021025,PhysRevA.89.021601}, but only a few studies have investigated the many-body properties of interacting fermions near a p-wave resonance\;\cite{PhysRevLett.90.053201,PhysRevA.70.030702,PhysRevA.71.045601,PhysRevA.88.012710,PhysRevA.98.020702,PhysRevLett.120.133401,PhysRevA.99.052704,PhysRevA.101.062702}.
For higher partial-wave resonances, challenges arise due to their short lifetimes and narrow resonance widths\;\cite{PhysRevLett.119.203402}. Recently, two broad d-wave resonances were observed via atom loss in a $^{85}$Rb\,-$^{87}$Rb mixture\;\cite{PhysRevLett.119.203402}. Very recently, a d-wave shape resonance and Feshbach resonance were observed in degenerate $^{41}$K and $^{39}$K gases, respectively\;\cite{Yao2019,PhysRevA.99.022701}. Moreover, both p-wave and d-wave Feshbach resonances were observed in the $^{133}$Cs\,-$^{6}$Li system\;\cite{Zhu2019}. This experimental progress has provided a platform to study the universal properties of few-body physics with d-wave resonant interactions.

In this paper, we investigate the TBR process in a heteronuclear system with d-wave resonant interactions. In real ultracold atomic systems, inter- and intraspecies interactions are generally not controlled independently. Thus, complications arise in heteronuclear three-body systems due to the two different scattering lengths. Near the interspecies Feshbach resonance, two identical atoms interact with each other over a small scattering length. Thus, we also focus on the more realistic case in which two heteronuclear atoms are in d-wave resonance while homonuclear atoms interact over a finite scattering length. As a result, ultracold gases of heteronuclear systems are expected to show rich few-body physics compared to the homonuclear case.

The TBR rates are obtained using quantum calculations based on a combination of the slow variable discretization (SVD) method, traditional hyperspherical adiabatic method and
$R$-matrix propagation method\;\cite{RevModPhys.68.1015,PhysRevA.65.042725,
PhysRevA.78.062701,PhysRevA.84.052721,Tolstikhin_1996,BALUJA1982299}.
Following the method of Ref\;\cite{PhysRevA.84.052721}, first, we solve the Schr$\mathrm{\ddot{o}}$dinger equation with the hyperradius divided into two regimes.
At short distances, the SVD method is employed to overcome the numerical difficulties at sharp nonadiabatic avoided crossings, and at large distances, the traditional adiabatic hyperspherical method is utilized to avoid the large memory and central processing unit (CPU) time needed in SVD.
Second, by propagating the $\underline{\mathcal{R}}$ matrix from short distances to large distances, we can obtain scattering properties through the $\underline{\mathcal{S}}$ matrix by matching the $\underline{\mathcal{R}}$ matrix with asymptotic functions and boundary conditions. The Lennard-Jones potential, which has been shown to be an excellent model potential, is utilized to mimic the interactions between atoms\;\cite{yujunwangPRl2012,Ulmanis2016cslia+,PhysRevA.86.062511,jiawangPRL2012, PhysRevA.90.022106}.

This paper is organized as follows:
In Sec. II, our calculation method and all necessary formulas for calculations are presented.
In Sec. III, we discuss the results and emphasize the significant role of intraspecies interactions in heteronuclear systems.
We then provide a brief summary.
Atomic units are applied throughout the paper unless stated otherwise.

\section{Theoretical formalism}
This numerical study focuses on the heteronuclear system with total angular momentum $J=0$. We use $m_i$ (i=1,2,3) to represent the mass of three atoms and use $r_{ij}$ to represent their distance. We choose the $^{85}$Rb\,-$^{87}$Rb\,-$^{87}$Rb system as an example and use the mass of Rb atoms in our calculations. In the center-of-mass frame, six coordinates are needed to describe the three-particle system.
Three of these coordinates are taken to be the Euler angles---$\alpha$, $\beta$, and $\gamma$---which specify the orientation of the body-fixed frame relative to the space-fixed frame.
The remaining degrees of freedom can be represented by hyperradius $R$ and the two hyperangles $\theta$ and $\phi$.
In our method, we employ Delves' hyperspherical coordinates.
We introduce the mass-scaled Jacobi coordinates. $\vec{\rho}_1$ is the vector from atom 1 to atom 2, with the reduced mass denoted by $\mu_1$; the second Jacobi $\vec{\rho}_2$ is measured from the diatom center of mass to the third atom, with reduced mass $\mu_2$. $\theta$ is the angle between $\vec{\rho}_1$ and $\vec{\rho}_2$. The hyperradius $R$ and hyperangle $\phi$ are defined as
\begin{equation}
\label{1}
\mu R=\mu_1 \rho^2_1+\mu_2 \rho^2_2
\end{equation}
and
\begin{equation}
\label{2}
\tan \phi =\sqrt{\frac{\mu_2}{\mu_1}}\frac{\rho_2}{\rho_1},\;\; 0 \leq \phi\leq\frac{\pi}{2},
\end{equation}
respectively, where $\mu$ is an arbitrary scaling factor that is chosen as $\mu=\sqrt{\mu_1\mu_2}$ in our calculations.
$R$ is the only coordinate with the dimension of length, which represents the size of the three-body system.
$\theta$, $\phi$ and the three Euler angles $(\alpha,\beta,\gamma)$ can be collectively represented by $\Omega$ $[\Omega\equiv(\theta,\phi,\alpha,\beta,\gamma)]$, which describe the rotation of the plane that contains the three particles.

In hyperspherical coordinates, the Schr$\mathrm{\ddot{o}}$dinger equation can be written in terms of the rescaled wave function $\psi_{\upsilon'}(R;\Omega)=\Psi_{\upsilon'}(R;\Omega)R^{5/2}\sin\phi\cos\phi$:
%\begin{widetext}
\begin{equation}
\label{3}
\bigg[-\frac{1}{2\mu}\frac{d^2}{dR^2}+\bigg(\frac{\Lambda^2-\frac{1}{4}}{2\mu R^2}
+V(R;\theta,\phi)\bigg)\bigg]\psi_{\upsilon'}(R;\Omega) =E\psi_{\upsilon'}(R;\Omega)\,,
\end{equation}
%\end{widetext}
where $\Lambda^2$ is the squared ``grand angular momentum operator``, whose expression is given in Ref\;\cite{CDLIN1995}.
The volume element relevant to integrals over $|\psi_{\upsilon'}(R;\Omega)|^2$ then becomes $dR \sin\theta d \theta d \phi d\alpha \sin \beta d\beta d\gamma$.
The index $\upsilon'$ labels the different independent solutions. The three-body interaction $V(R;\theta,\phi)$ in Eq.\;(\ref{3}) is taken to be a sum of the three pairwise two-body interactions $\upsilon(r_{ij})$:
\begin{equation}
\label{4}
V(R;\theta,\phi) = \upsilon(r_{12})+\upsilon(r_{13})+\upsilon(r_{23})\,.
\end{equation}
The interparticle distances $r_{ij}$ can be described in terms of the internal coordinates as follows:
\begin{align}
\label{5}
&r_{12}=R\sqrt{\frac{\mu}{\mu_1}}\cos\phi, \\
&r_{23}=R\Big(\frac{\mu}{\mu_2}\sin^2\phi
+\frac{1}{4}\frac{\mu}{\mu_1}\cos^2\phi
-\frac{1}{2}\sin 2\phi\cos\theta\Big)^{1/2},\\
&r_{13}=R\Big(\frac{\mu}{\mu_2}\sin^2\phi
+\frac{1}{4}\frac{\mu}{\mu_1}\cos^2\phi
+\frac{1}{2}\sin 2\phi\cos\theta\Big)^{1/2}.
\end{align}

The wave function $\psi_{\upsilon'}$ can then be expanded with the complete, orthonormal adiabatic channel functions $ \Phi_\upsilon$ as
\begin{align}
\label{8}
 \psi_{\upsilon'}(R;\Omega)=\sum^\infty_{\nu=0}F_{\nu \upsilon'}(R)\Phi_\nu(R;\Omega)\,.
\end{align}
We determine the adiabatic potentials $U_{\nu}(R)$ and corresponding channel
functions $\Phi_{\nu}(R;\Omega)$ at a fixed $R$ by solving the following adiabatic eigenvalue equation:
\begin{equation}
\label{9}
\bigg(\frac{\Lambda^{2}-\frac{1}{4}}{2\mu R^{2}}+
V(R;\theta,\phi)  \bigg)\Phi_{\nu}(R;\Omega)=U_{\nu}(R)\Phi_{\nu}(R;\Omega)\,.
\end{equation}
The channel function is further expanded on Wigner rotation matrices $D^J_{KM}$ as
\begin{align}
\label{10}
\Phi_\nu^{J\Pi M}(R;\Omega)=\sum^J_{K=0}u_{\nu K}(R;\theta,\phi)\overline{D}^{J\Pi}_{KM}(\alpha,\beta,\gamma),
\end{align}
\begin{align}
\label{11}
\overline{D}^{J\Pi}_{KM}=\frac{1}{4\pi}\sqrt{2J+1}
\left[D^J_{KM}+(-1)^{K+J}\Pi D^J_{-KM}\right],
\end{align}
where $J$ is the total nuclear orbital angular momentum, $M$ is its projection onto the laboratory-fixed axis, and $\Pi$ is the parity with respect to the inversion of the nuclear coordinates. The quantum number $K$ denotes the projection of $J$ onto the body-frame $z$ axis and takes the values $K=J,J-2,\ldots,-(J-2),-J$ for the ``parity-favored'' case, $\Pi=(-1)^J$, and $K=J-1,J-3,\ldots,-(J-3),-(J-1)$ for the ``parity-unfavored'' case, $\Pi=(-1)^{J+1}$.

For the $^{85}$Rb\,-$^{87}$Rb\,-$^{87}$Rb system, the wave function is symmetric with respect to the exchange of the two $^{87}$Rb atoms, and thus, this exchange symmetry can be built into the boundary conditions of the body-frame components as follows:
\begin{align}
\label{12}
&P_{12}\overline{D}^{J\Pi}_{KM}=\Pi(-1)^K\overline{D}^{J\Pi}_{KM},\\
&P_{12}\theta=\pi-\theta.
\end{align}
For even parity, $u_{\nu K}$ should be symmetric about $\pi/2$ with even $K$ and antisymmetric with odd $K$. For odd parity, $u_{\nu K}$ should be antisymmetric for even $K$ and symmetric for odd $K$. To satisfy the permutation requirements, $u_{\nu K}$ is expanded with symmetric B-spline basis sets
\begin{align}
\label{14}
u_{\nu K}(R;\theta,\phi)=\sum^{N_{\phi}}_i\sum^{N_{\theta}/2}_j
c_{i,j}B_i(\phi)\left(B_j(\theta)+B_{N_{\theta}+1-j}(\pi-\theta)\right),
\end{align}
or antisymmetric B-spline basis sets
\begin{align}
\label{15}
u_{\nu K}(R;\theta,\phi)=\sum^{N_{\phi}}_i\sum^{N_{\theta}/2}_j
c_{i,j}B_i(\phi)\left(B_j(\theta)-B_{N_{\theta}+1-j}(\pi-\theta)\right),
\end{align}
where $N_{\theta}$ and $N_{\phi}$ are the sizes of the basis sets in the $\theta$ direction and $\phi$ direction, respectively.
The constructed symmetric B-spline basis sets utilized in the $\theta$ direction reduce the number of basis functions to $N_{\theta}/2$.

The goal of our scattering study is to determine the scattering matrix $\underline{\mathcal{S}}$ from the solutions of Eq.\;(\ref{3}). We calculate the $\underline{\mathcal{R}}$ matrix, which is defined as
\begin{align}
\label{16}
\underline{\mathcal{R}}(R)=\underline{\textsl{F}}(R)[\widetilde{\underline{\textsl{F}}}(R)]^{-1}\,,
\end{align}
where matrices $\underline{\textsl{F}}$ and $\widetilde{\underline{\textsl{F}}}$ can be calculated from the solution of Eq.~(\ref{3}) and Eq.~(\ref{9}) by
\begin{align}
\label{17}
F_{\nu,\upsilon'}(R)=\int d\Omega \Phi_{\nu}(R;\Omega)^{*}\psi_{\upsilon'}(R;\Omega)\,,
\end{align}
\begin{align}
\label{18}
\widetilde{F}_{\nu,\upsilon'}(R)=\int d\Omega \Phi_{\nu}(R;\Omega)^{*}\frac{\partial}{\partial R}\psi_{\upsilon'}(R;\Omega)\,.
\end{align}

Following the method of Ref.\;\cite{PhysRevA.84.052721}, we divide the hyperradius into $(N-1)$ intervals with the set of grid points $R_1<R_2<\cdots R_N$. At a short distance, we use the SVD method to solve Eq.\;(\ref{3}) in the interval $[R_i,R_i+1]$. In the SVD method, the total wave function $\psi_{\upsilon'}(R;\Omega)$ is expanded in terms of the discrete variable representation (DVR) basis $\pi_i$ and the channel functions $\Phi_{\nu}(R;\Omega)$ as
%Editor: Please ensure that the intended meaning has been maintained in the above edit.
\begin{align}
\label{19}
 \psi_{\upsilon'}(R;\Omega)=\sum^{N_{DVR}}_i\sum^{N_{chan}}_{\nu}C^{ \upsilon'}_{i\nu}\pi_i(R)\Phi_\nu(R_i;\Omega)\,,
\end{align}
where $N_{DVR}$ is the number of DVR basis functions and $N_{chan}$ is the number of included channel functions.
%Editor: Please ensure that the intended meaning has been maintained in the above edit.
Inserting $\psi_{\upsilon'}(R;\Omega)$ into the three-body Schrodinger equation yields the standard algebraic problem for the coefficients $C^{\upsilon'}_{i\nu}$:
\begin{align}
\label{20}
 \sum^{N_{DVR}}_{j}\sum^{N_{chan}}_{\mu}\mathcal{T}_{ij}
 \mathcal{O}_{i\nu,j\mu}C^{\upsilon'}_{i\nu}+U_\nu (R_i)C^{\upsilon'}_{i\nu}=E^{\upsilon'}C^{\upsilon'}_{i\nu},
\end{align}
where
\begin{align}
\label{21}
\mathcal{T}_{ij}=\frac{1}{2\mu}\int^{R_{i+1}}_{R_{i}}\frac{d}{dR}\pi_i(R)\frac{d}{dR}\pi_j(R)dR
\end{align}
are the kinetic energy matrix elements; ${R_{i}}$ and ${R_{i+1}}$ are the boundaries of the calculation box; and
\begin{align}
\label{22}
\mathcal{O}_{i\nu,j\mu}=\langle\Phi_\nu(R_i;\Omega)|\Phi_\mu(R_j;\Omega)\rangle
\end{align}
are the overlap matrix elements between the adiabatic channels defined at different quadrature points.

At large distances, the traditional adiabatic hyperspherical method is used to solve Eq.\;(\ref{3}).
When substituting the wave functions $\psi(R;\Omega)$ into Eq.\;(\ref{3}), a set of coupled ordinary differential equations is obtained:
\begin{align}
\label{28}
[-\frac{1}{2\mu}\frac{d^2}{dR^2}+U_\nu(R)- E]F_{\nu,\upsilon'}(R)
-\frac{1}{2\mu}\sum_{\mu}[2P_{\mu\nu}(R)\frac{d}{dR}+Q_{\mu\nu}(R)]F_{\mu \upsilon'}(R)=0\,,
\end{align}
where
\begin{align}
\label{29}
P_{\mu\nu}(R)=\int d\Omega \Phi_{\mu}(R;\Omega)^{*}\frac{\partial}{\partial R}\Phi_{\nu}(R;\Omega)
\end{align}
and
\begin{align}
\label{30}
Q_{\mu\nu}(R) = \int d\Omega \Phi_{\mu}(R;\Omega)^{*}\frac{\partial^{2}}{\partial R^{2}}\Phi_{\nu}(R;\Omega)
\end{align}
are the nonadiabatic couplings that control the inelastic transitions and the width of the resonance supported by adiabatic potential
$U_\nu(R)$ .
In our calculations, the relation between $\underline{P}$ and $\underline{Q}$ is $\frac{d}{d R} \underline{P}=-\underline{P^{2}}+\underline{Q}$, where
\begin{align}
\label{31}
P_{\nu \mu}^{2}(R)=-\int d \Omega \frac{\partial}{\partial R} \Phi_{\nu}(R ; \Omega)^{*} \frac{\partial}{\partial R} \Phi_{\mu}(R ; \Omega).
\end{align}
The coupling matrices have the following properties:
$P_{\nu \mu}=-P_{\mu \nu}$ and $P_{\nu \mu}^{2}=P_{\mu \nu}^{2}$, which leads to $P_{\nu \nu}=0,$ and $Q_{\nu \nu}=-P_{\nu \nu}^{2}$.
The effective hyperradial potentials, which include hyperradial kinetic energy contributions with the $P_{\nu\nu}^2$ term, are more physical than adiabatic hyperpotentials and are defined as
\begin{align}
\label{32}
W_{\nu \nu}(R)=U_{\nu}(R)-\frac{\hbar^{2}}{2 \mu} P_{\nu \nu}^{2}(R).
\end{align}

Next, the $R$-matrix propagation method is employed. Over the interval $[R_{1},R_{2}]$, for a given $\underline{\mathcal{R}}$ matrix Eq.\;(\ref{16}), the $R$-matrix propagation method can be used to calculate the corresponding $\underline{\mathcal{R}}$ matrix at another point $R = R_{2}$ as follows:

\begin{align}
\label{33}
\underline{\mathcal{R}}(R_2)=\underline{\mathcal{R}}_{22}
-\underline{\mathcal{R}}_{21}\left[\underline{\mathcal{R}}_{11}
+\underline{\mathcal{R}}(R_1)\right]^{-1}\underline{\mathcal{R}}_{12}.
\end{align}

The $\underline{\mathcal{K}}$ matrix can be expressed in the following matrix equation:
\begin{align}
\label{34}
\underline{\mathcal{K}}=
(\underline{f}-\underline{f}'\underline{\mathcal{R}})
(\underline{g}-\underline{g}'\mathcal{R})^{-1}\,,
\end{align}
where $f_{\nu \nu'}=\sqrt{\frac{2\mu }{\pi k_\nu}} k_\nu R j_{l_\nu}(k_\nu R)\delta_{\nu \nu'}$ and $g_{\nu \nu'}=\sqrt{\frac{2\mu }{\pi k_\nu}} k_\nu R n_{l_\nu}(k_\nu R)\delta_{\nu \nu'}$ are the diagonal matrices of energy-normalized spherical Bessel and Neumann functions.
For the recombination channel, $l_\nu$ is the angular momentum of the third atom relative to the dimer, and $k_\nu$ is given by $k_{\nu}=\sqrt{2 \mu\left(E-E_{2 b}\right)}$.
For the entrance channel, $l_{\nu}=\lambda_{\nu}+3 / 2$, and $k_{\nu}=\sqrt{2 \mu E}$.
The scattering matrix $\underline{\mathcal{S}}$ is related to $\underline{\mathcal{K}}$ as follows:
\begin{align}
\label{35}
\underline{\mathcal{S}}=(\underline{1}+i\underline{\mathcal{K}})(\underline{1}-i\underline{\mathcal{K}})^{-1}\,.
\end{align}

Using the convention of Mott and Massey\;\cite{PhysRevLett.103.153201}, the N-body cross-section in d dimensions is defined as
\begin{align}
\label{36}
\sigma_{fi}(J^\Pi)=N_p\bigg(\frac{2\pi}{k_i}\bigg)^{d-1}\frac{1}{\Omega(d)}\sum_i(2J+1)|S^{J\Pi}_{fi}-\delta_{fi}|^2\,,
\end{align}
where $\Omega(d)=2\pi^{d/2}/\Gamma(d/2)$ is the total solid angle in d dimensions and $N_p$ is the number of terms in the permutation symmetry projection operator.
In the $^{85}$Rb\,-$^{87}$Rb\,-$^{87}$Rb system, $N_p=2\,!$, $d=6$, and the total TBR rate is then
\begin{align}
\label{37}
K_3=\frac{k}{\mu}\sigma_3=\sum_{J,\Pi}K_3^{J,\Pi}=2!\sum_{J,\Pi}\sum_{f,i}\frac{32(2J+1)\pi^2}{\mu k^4}|S^{J,\Pi}_{i\rightarrow f}|^2,
\end{align}
where i and f label the three-body continuum (incident) channel and TBR (outgoing) channel, respectively.
$\sigma_3$ is the generalized TBR cross-section.
$K_3^{J,\Pi}$ is the partial recombination rate corresponding to $J^\Pi$ symmetry, and $k=(2\mu E)^{1/2}$ is the wave number in the incident channels.

Since experiments are performed at a fixed temperature instead of a fixed energy, the thermal average becomes crucial for proper comparison with the experiment.
Assuming a Boltzmann distribution, the thermally averaged recombination rates are given by
\begin{align}
\label{38}
\langle K_3\rangle(T)=\frac{\int K_3(E)E^2e^{-E/(k_BT)}dE}{\int E^2e^{-E/(k_BT)}dE}=\frac{1}{2(k_BT)^3}\int K_3(E)E^2e^{-E/(k_BT)}dE.
\end{align}
The results presented in \ref{subsectionC} are given for $T=120~nK$.

In our calculations, Eq.\;(\ref{9}) is solved with 134 SVD sectors and 10 SVD points in each sector for $R<2\,000 a_0$. In the interval $2\,000 a_0 < R < 22\,000 a_0$, we use the traditional adiabatic hyperspherical method with $P_{\mu\nu}$ and $Q_{\mu\nu}$ calculated by an improved method in Ref\;\cite{PhysRevA.86.062511}.
The matrix elements of coupling $P_{\mu\nu}$ and effective potential $W_{\nu\nu} $ can be fitted to an inverse polynomial series at a large distance, and the fitting results of $P_{\mu\nu}$ and $W_\nu$ are applied beyond $R =22\,000 a_0$.

\section{Results and discussion}

\subsection{Hyperspherical potential curves near the interspecies d-wave resonance}

\label{subsectionA}
We use the Lennard-Jones potential to model the interactions between two atoms. The advantages of the Lennard-Jones potential is that it has the van der Waals length and can also avoid numerical difficulties in a short range.\;\cite{jiawangPRL2012}. Thus, it has been widely used and
proven
 to be an excellent model potential to explore van der Waals universality in Efimov physics \;\cite{yujunwangPRl2012,Ulmanis2016cslia+,PhysRevA.86.062511,jiawangPRL2012, PhysRevA.90.022106}. The potential is expressed in the following form:
\begin{equation}
\label{39}
\upsilon(r_{ij})= -\frac{C_{6,ij}}{r^6_{ij}}\bigg[1-\frac{1}{2}(\frac{\gamma_{ij}}{r_{ij}})^6\bigg]\,.
\end{equation}

In this study, $\gamma_{ij}$ is adjusted to give the desired scattering length and number of bound states.
The value of $C_6$ for two Rb atoms we adopted here is 4698 from Ref.\;\cite{PhysRevA.89.022703}.
The low-energy behavior of the $l$ th partial wave phase shift for scattering by a long-range central potential $1/r^s (s>2)$ satisfies
\begin{align}
\label{40}
\tan \delta_{l}(k, \infty) \sim-k^{2 l+1} \lambda_{l}-\frac{\pi}{2^{s}} \frac{\Gamma(s-1) \Gamma\left(l+\frac{3}{2}-\frac{1}{2} s\right)}{\Gamma^{2}\left(\frac{1}{2} s\right) \Gamma\left(l+\frac{1}{2}+\frac{1}{2} s\right)}2\mu_{2b} C_{s} k^{s-2},
\end{align}
with $2<s<2 l+3$\;\cite{doi:10.1063/1.1703889,PhysRevA.74.052715,PhysRevA.87.063629}. For the Lennard-Jones potential $s=6$, the scattering phase shift of the d-wave has the following expansion:
\begin{align}
\label{41}
\tan \delta_{2}(k, \infty) \sim-k^5 \lambda_{2}-\frac{\pi}{2^6} \frac{\Gamma(5) \Gamma\left(2\right)}{\Gamma^{2}\left(3\right) \Gamma\left(11/2\right)}2\mu_{2b} C_{6} k^4.
\end{align}
$a_d=\lambda_2^{1/5}$ is denoted as the ``d-wave scattering length'' to characterize the findings in terms of the d-wave interactions, which diverges when a d-wave dimer is almost bound.

\begin{figure}
 \centering
   \includegraphics[width=0.7\linewidth]{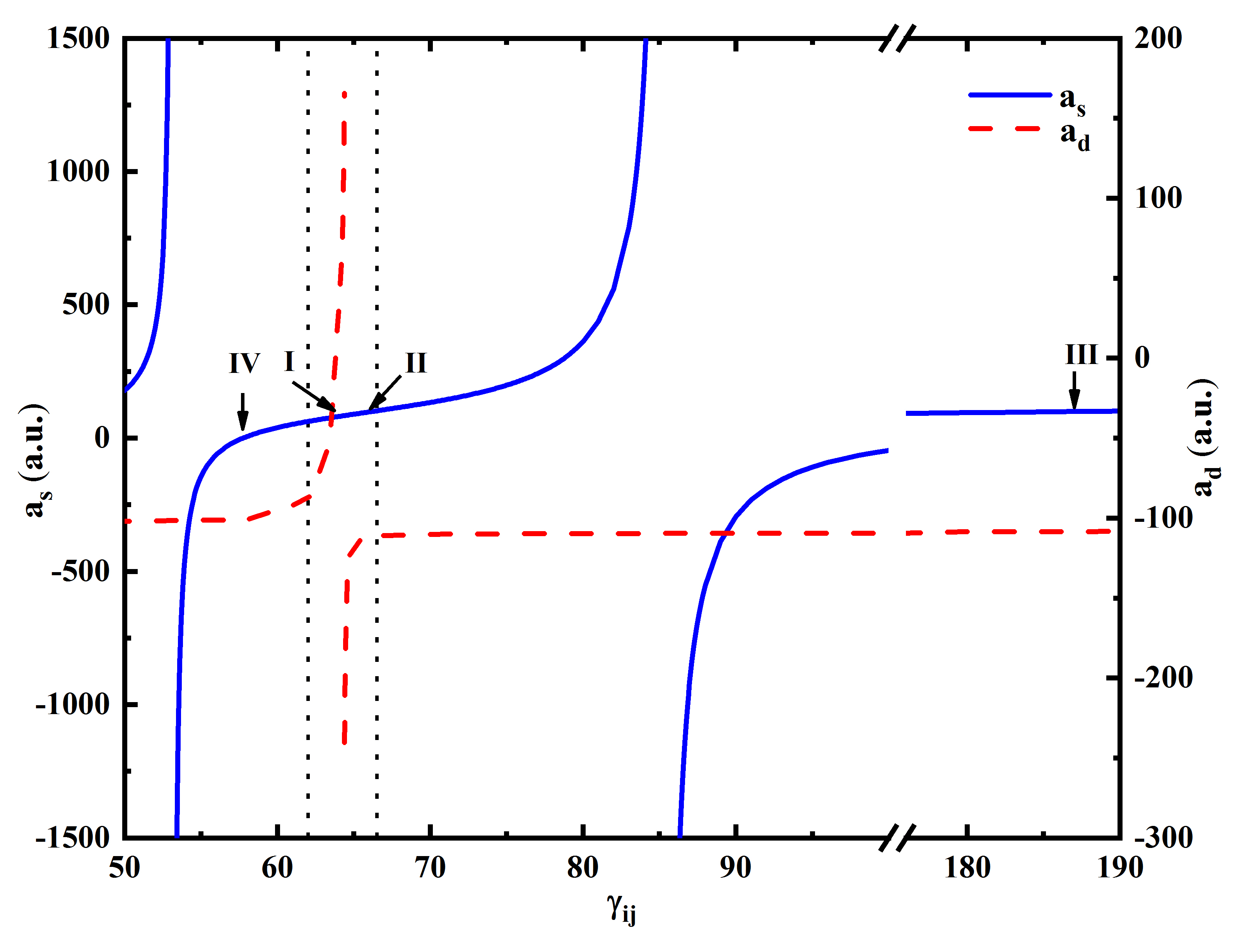}
\caption{(Color online) Two-body s-wave scattering length (blue solid line) and d-wave scattering length (red dashed line) as a function of the adjusting parameter $\gamma_{ij}$. The $^{85}$Rb\,-$^{87}$Rb interaction range is adjusted around the two-body d-wave resonance that is shown between the two black dotted lines. For the $^{87}$Rb\,-$^{87}$Rb interaction, points I, II, III and IV
%Quality Control Editor 2: Please note that some text appears to be missing here. Please consider adding any missing information.
are focused on
. }
   \label{fig1}
\end{figure}

Figure\;\ref{fig1} shows the s-wave scattering length $a_s$ (blue solid line) and d-wave scattering length $a_d$ (red dashed line) as a function of $\gamma_{ij}$. No obvious difference is observed between $^{85}$Rb\,-$^{87}$Rb and $^{87}$Rb\,-$^{87}$Rb, so we show the results for only one case. The vertical dotted lines enclose the $^{85}$Rb\,-$^{87}$Rb parameter range considered in our numerical calculations. In this range, one two-body s-wave bound state exists before the d-wave bound state emerges. For a homonuclear interaction, we focus on the case in which two $^{87}$Rb atoms are in d-wave resonance (point I in Fig.\;\ref{fig1}) and the more realistic case, in which $^{87}$Rb\,-$^{87}$Rb interact via the s-wave scattering length $a_s=100a_0$, which is indicated by arrows II and III. As shown in Fig.\;\ref{fig1}, the two $^{87}$Rb atoms have one s-wave bound state at point II and IV. At point III, the two $^{87}$Rb atoms have no bound states and are located away from the d-wave resonance.

\begin{figure}[htbp]
\centering
\subfigure{
\includegraphics[width=7.5cm]{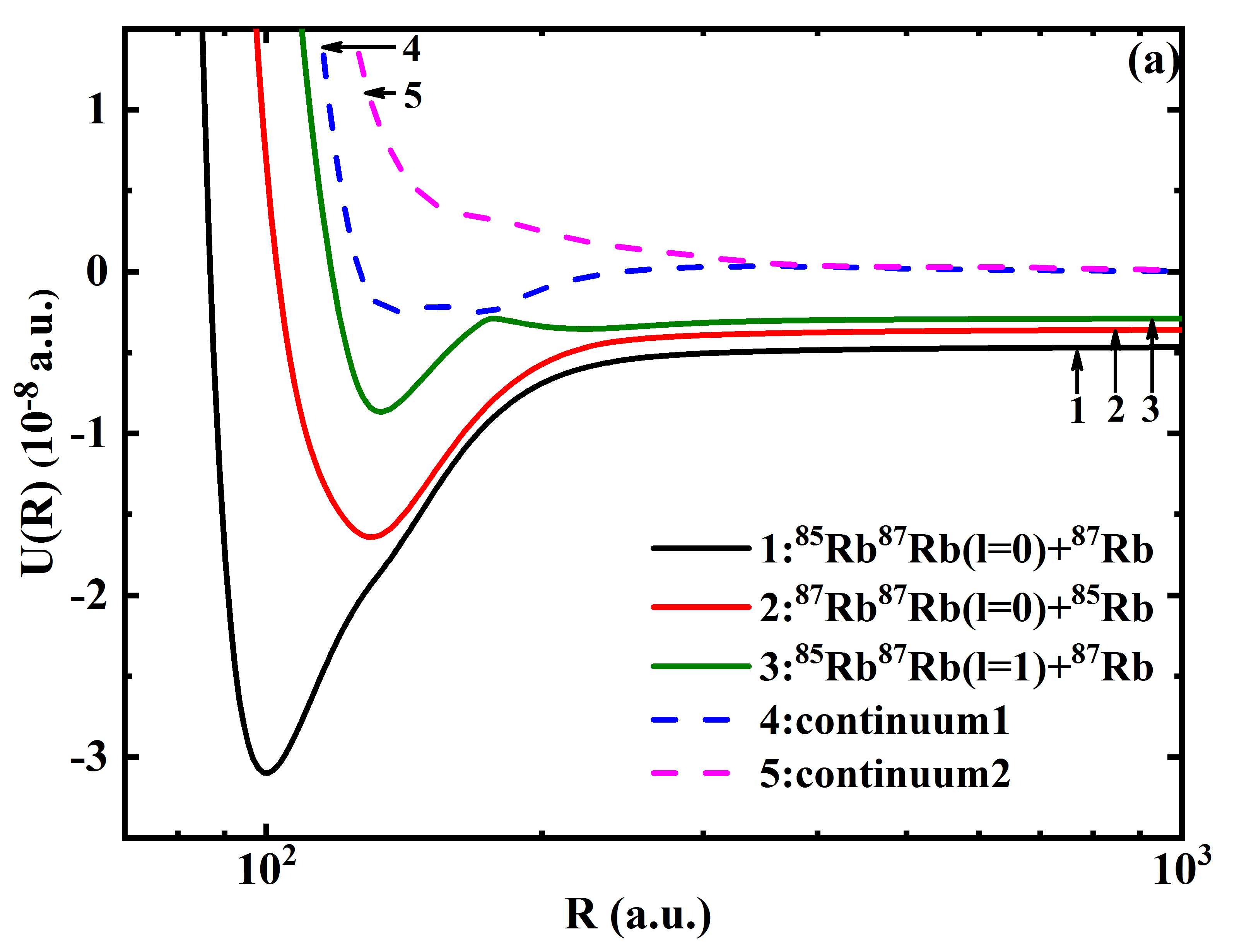}
%\caption{fig1}
\label{fig2a}
}
%\quad
\subfigure{
\includegraphics[width=7.5cm]{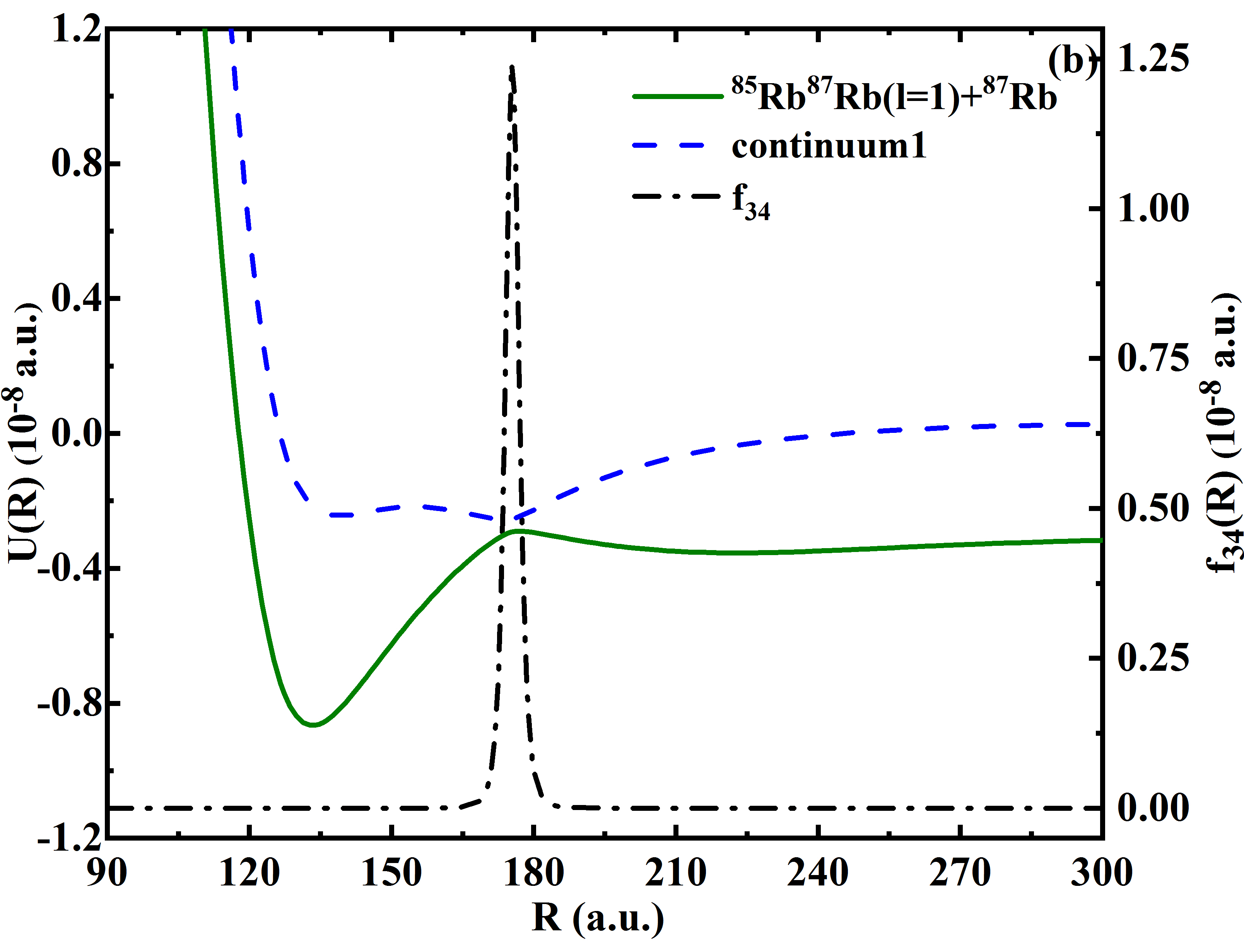}
\label{fig2b}
}
\subfigure{
\includegraphics[width=7.5cm]{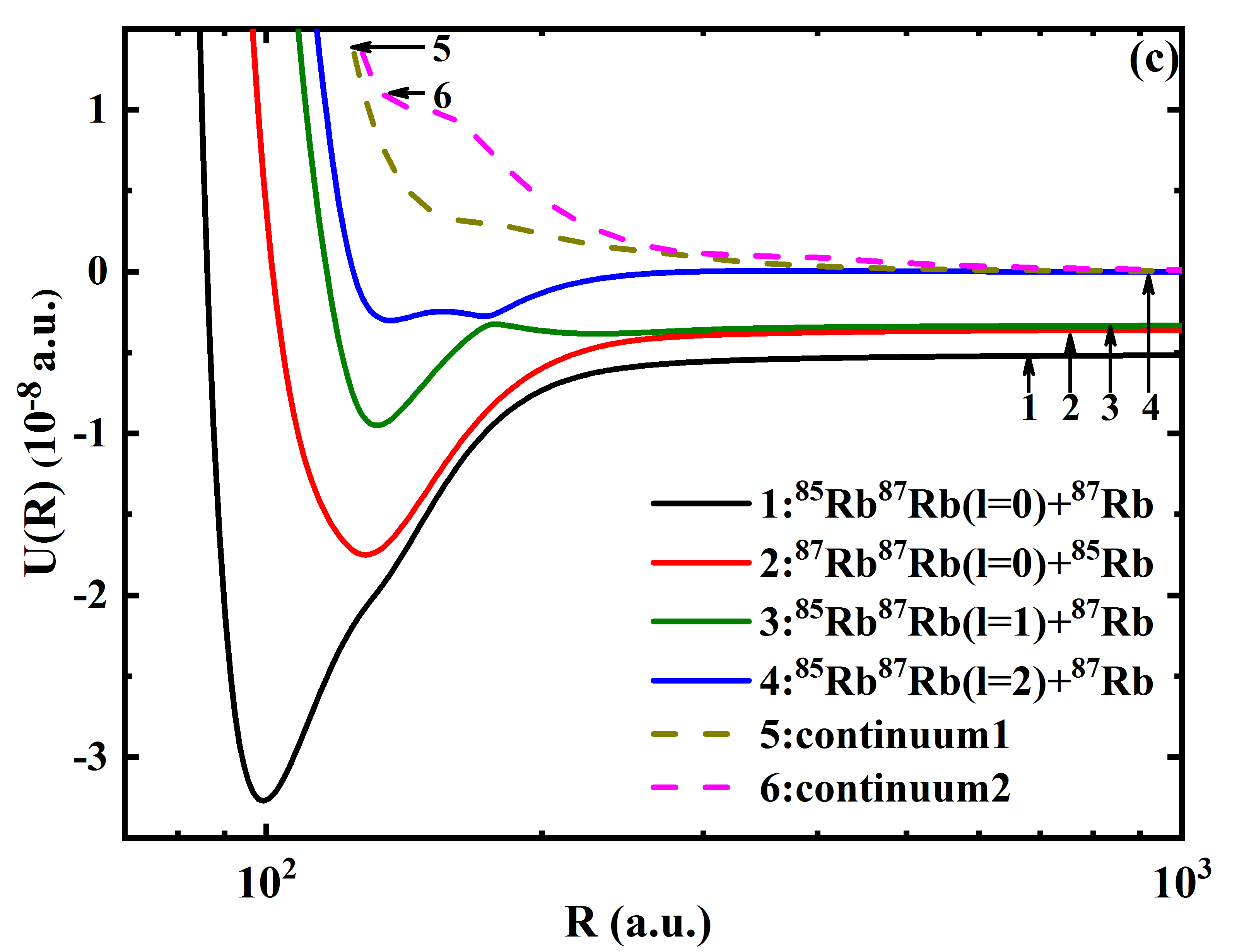}
%\caption{fig1}
\label{fig2c}
}
%\quad
\subfigure{
\includegraphics[width=7.5cm]{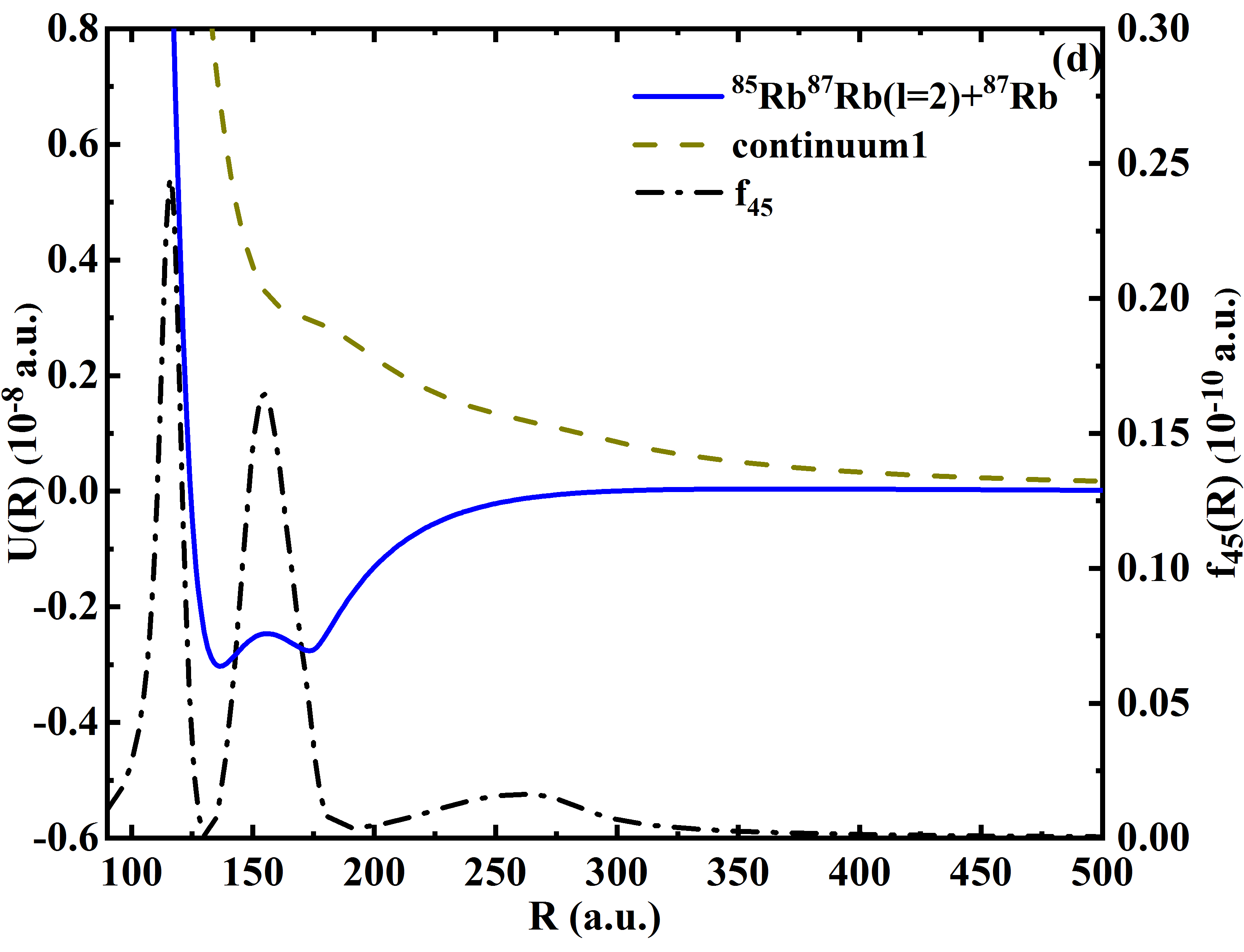}
\label{fig2d}
}
\caption{(Color online) Three-body adiabatic potential curves for the $^{85}$Rb\,-$^{87}$Rb-$^{87}$Rb system with $^{85}$Rb\,-$^{87}$Rb d-wave scattering length $a_d=-115a_0$ in (a) and $a_d=90a_0$ in (c). The $^{87}$Rb\,-$^{87}$Rb s-wave scattering length is $a_s=100a_0$ with the parameter $\gamma_{ij}$ adjusted to point II in Fig.\;\ref{fig1}. (b) and (d) show the avoided crossings between the highest recombination channel and the first entrance channel with their nonadiabatic coupling strength corresponding to (a) and (c), respectively.}
\label{fig2}
\end{figure}

In the scattering process, the adiabatic potential curves $U_\nu(R)$ are important in understanding three-body physics. Figure\;\ref{fig2} shows the hyperspherical potential curves of the $^{85}$Rb\,-$^{87}$Rb--$^{87}$Rb system with the $^{85}$Rb\,-$^{87}$Rb d-wave scattering length $a_{d} = -115 a_0$ (Fig.\;\ref{fig2a}) and $a_{d} = 90 a_0$ (Fig.\;\ref{fig2c}). $^{87}$Rb\,-$^{87}$Rb interact via the s-wave scattering length $a_s=100 a_0$ with the parameter $\gamma_{ij}$ adjusted to point II in Fig.\;\ref{fig1}. In each case, the solid potential curves correspond to the TBR channels and asymptotically approach the dimer binding energy. The effective potentials for these channels exhibit asymptotic behavior, given by
\begin{align}
\label{42}
W_f(R)=\frac{l_f(l_f+1)}{2 \mu R^2}+E^{(f)}_{2b},
\end{align}
where $E^{(f)}_{2b}$ is the dimer energy and $l_f$ is the relative orbital angular momentum between the atom and the dimer.
The subscript $f$ distinguishes the recombination channels. The dashed lines in Fig.\;\ref{fig2a} and Fig.\;\ref{fig2c} denote the three-body breakup channels (or entrance channels); i.e., all three atoms exist far from each other as $R\rightarrow \infty$, where the potentials behave as
\begin{align}
W_i(R)=\frac{\lambda_i(\lambda_i+4)+15/4}{2 \mu R^2}.
\label{43}
\end{align}
The values of $\lambda_i$ are nonnegative integers determined by $J^\Pi$ and the identical particle symmetry\;\cite{PhysRevA.65.010705}.
We use the dimensionless quantity of the nonadiabatic coupling strength defined by
\begin{align}
f_{v v^{\prime}}(R)=\frac{P_{v v^{\prime}}(R)^{2}}{2 \mu\left[U_{v}(R)-U_{v^{\prime}}(R)\right]}
\end{align}
to characterize the nonadiabatic coupling magnitude, which mainly controls the recombination process.
The coupling strength between the highest recombination channel and the lowest entrance channel is shown in Fig.\;\ref{fig2b} and Fig.\;\ref{fig2d} .

\begin{figure}[htbp]
\centering
\subfigure{
\includegraphics[width=7cm]{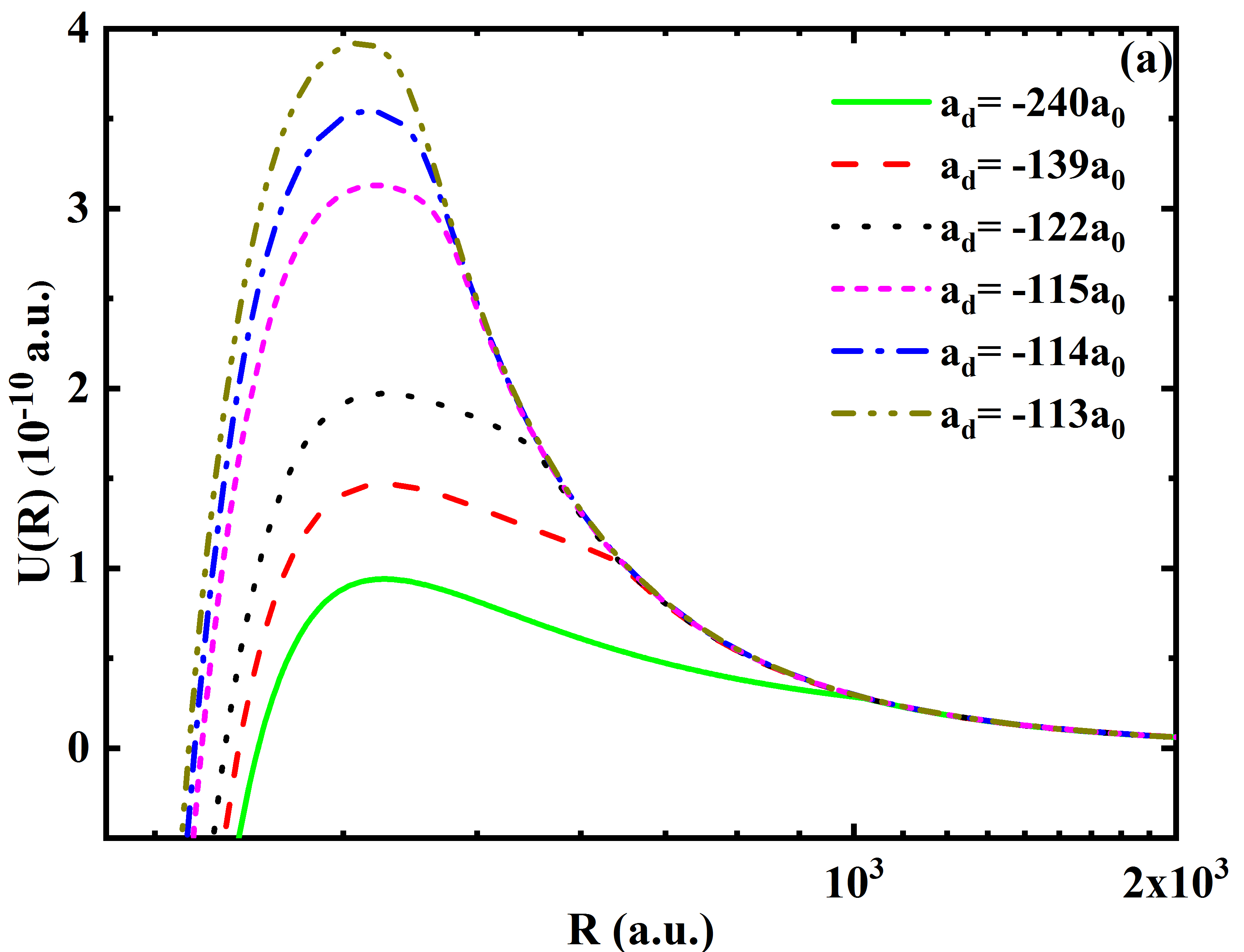}
\label{fig3a}
}
%\quad
\subfigure{
\includegraphics[width=7cm]{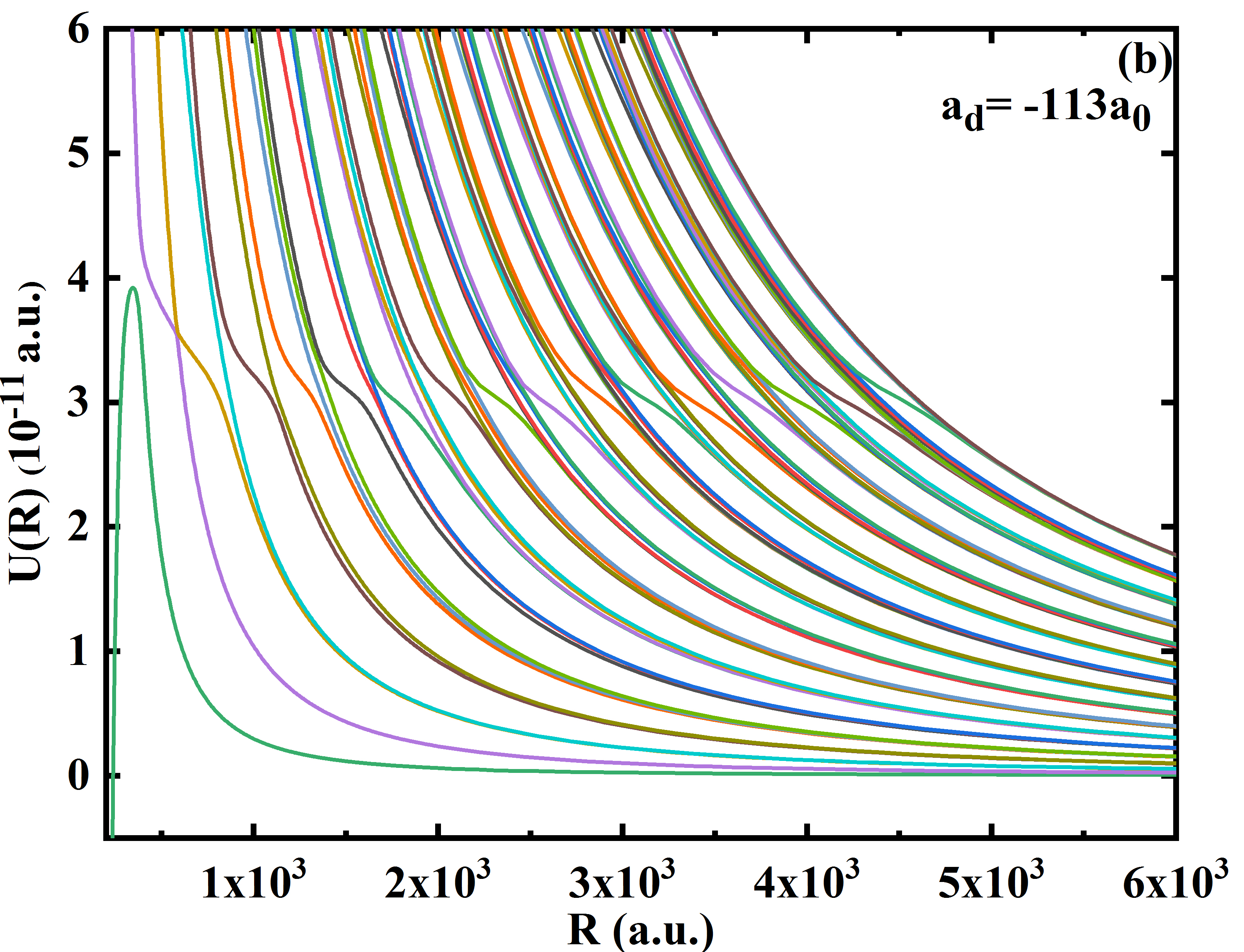}
\label{fig3b}
}
\subfigure{
\includegraphics[width=7cm]{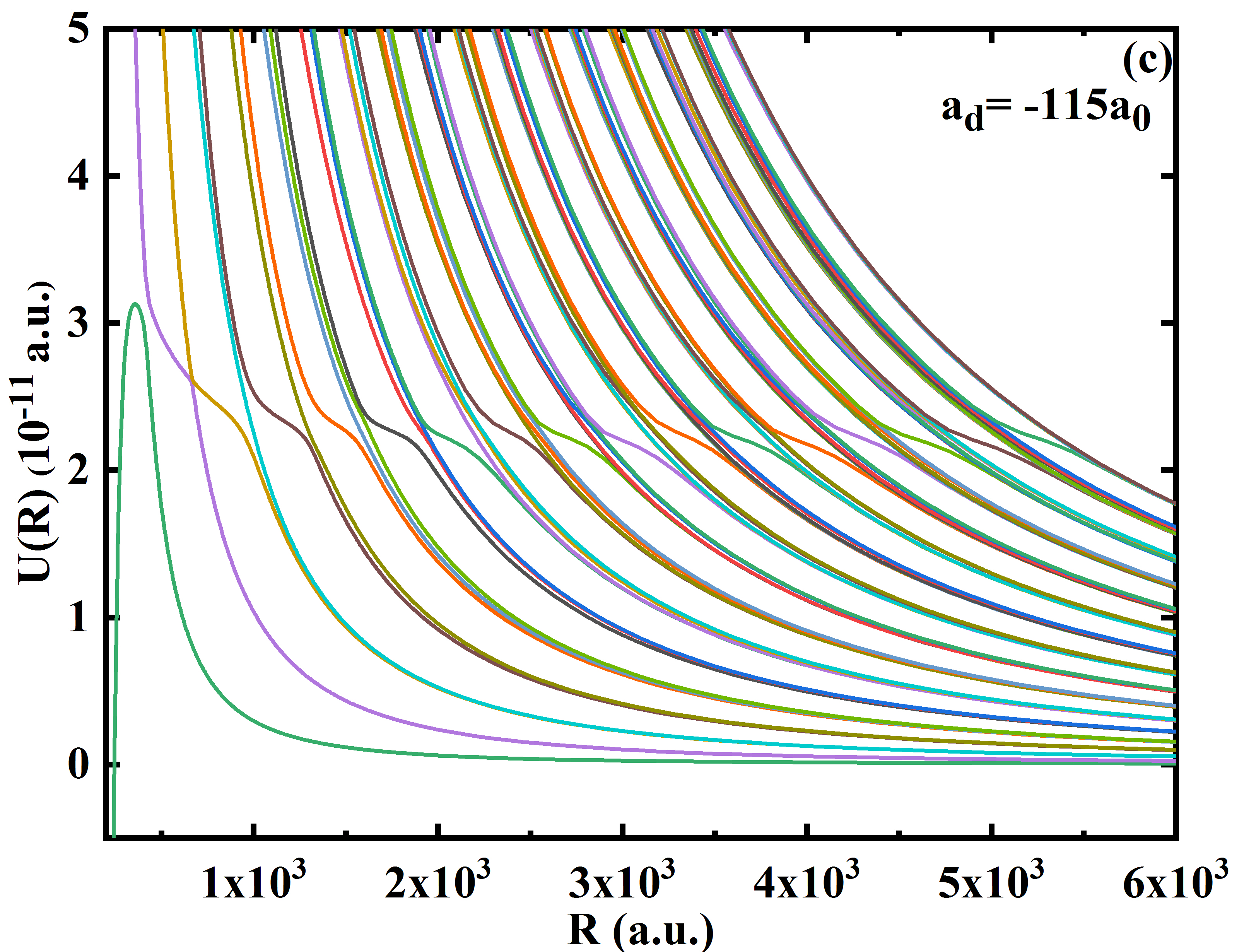}
\label{fig3c}
}
%\quad
\subfigure{
\includegraphics[width=7cm]{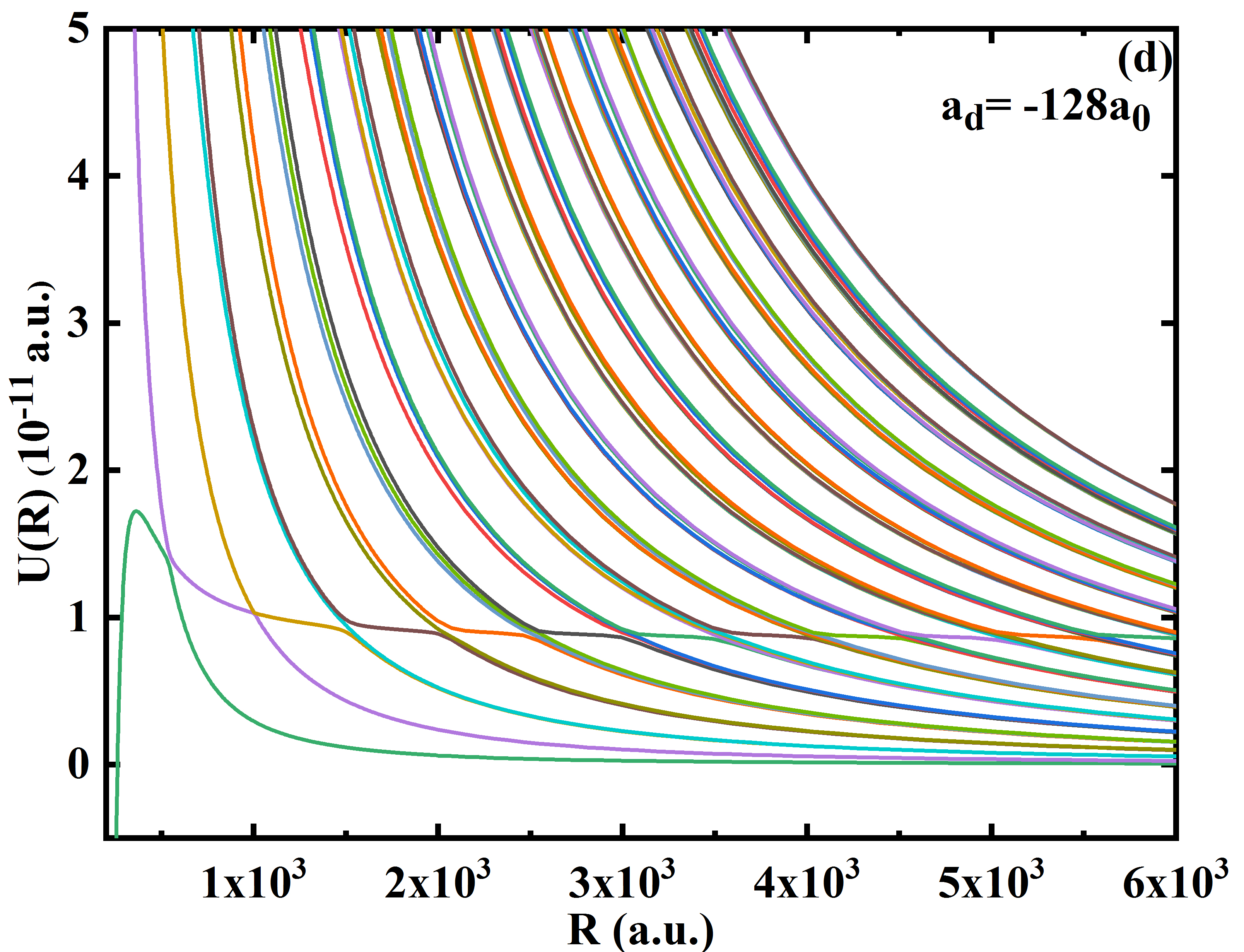}
\label{fig3d}
}
\caption{(Color online) (a) Lowest entrance channels with barriers in $^{85}$Rb\,-$^{87}$Rb-$^{87}$Rb for the different $^{85}$Rb\,-$^{87}$Rb d-wave scattering lengths; (b), (c) and (d) are potential curves that show a series of avoided crossings near the $^{85}$Rb\,-$^{87}$Rb two-body d-wave resonance. These avoided crossings approach the three-body breakup threshold when $a_d\rightarrow - \infty$. The $^{87}$Rb\,-$^{87}$Rb s-wave scattering length is $a_s=100a_0$ with the parameter $\gamma_{ij}$ adjusted to point II in Fig.\;\ref{fig1}.}
\label{fig3}
\end{figure}

When the $^{85}$Rb\,-$^{87}$Rb d-wave scattering length $a_d$ is negative, as shown in Fig.\;\ref{fig2a} and \ref{fig2b}, the nonadiabatic coupling strength between the lowest entrance channel and the first recombination channel is localized at a short distance. Here, recombination occurs primarily by tunneling through the potential barrier in the lowest three-body entrance channel to reach the region of large coupling. Thus, the potential barrier in the lowest three-body entrance channel has an important role in the recombination process. Figure\;\ref{fig3a} shows the barrier for several negative $^{85}$Rb\,-$^{87}$Rb scattering lengths $a_d$. The height of the barrier decreases as the $^{85}$Rb\,-$^{87}$Rb d-wave interaction strengthens. Diminishing of the potential barrier in the entrance channel is responsible for the recombination enhancement on this scattering side\;\cite{Kartavtsev2009}. In addition, when the $^{85}$Rb\,-$^{87}$Rb interaction becomes sufficiently strong, a two-body d-wave shape resonance appears\;\cite{DIncao2009}. As a result, this resonance will produce a series of avoided crossings in the three-body potential curves at energies near the position of the resonance, as shown in Figs.\;\ref{fig3b}-\ref{fig3d}. Ref.\;\cite{Suno_2003} also found this phenomenon in a three-fermion system when the two-body scattering volume was negative. The positions of these avoided crossings might be expected to approach the three-body breakup threshold when the $^{85}$Rb\,-$^{87}$Rb d-wave interaction strengthens. This result can be demonstrated by Figs.\;\ref{fig3b}-\ref{fig3d}.

\begin{figure}[htbp]
\centering
\subfigure{
\includegraphics[width=7cm]{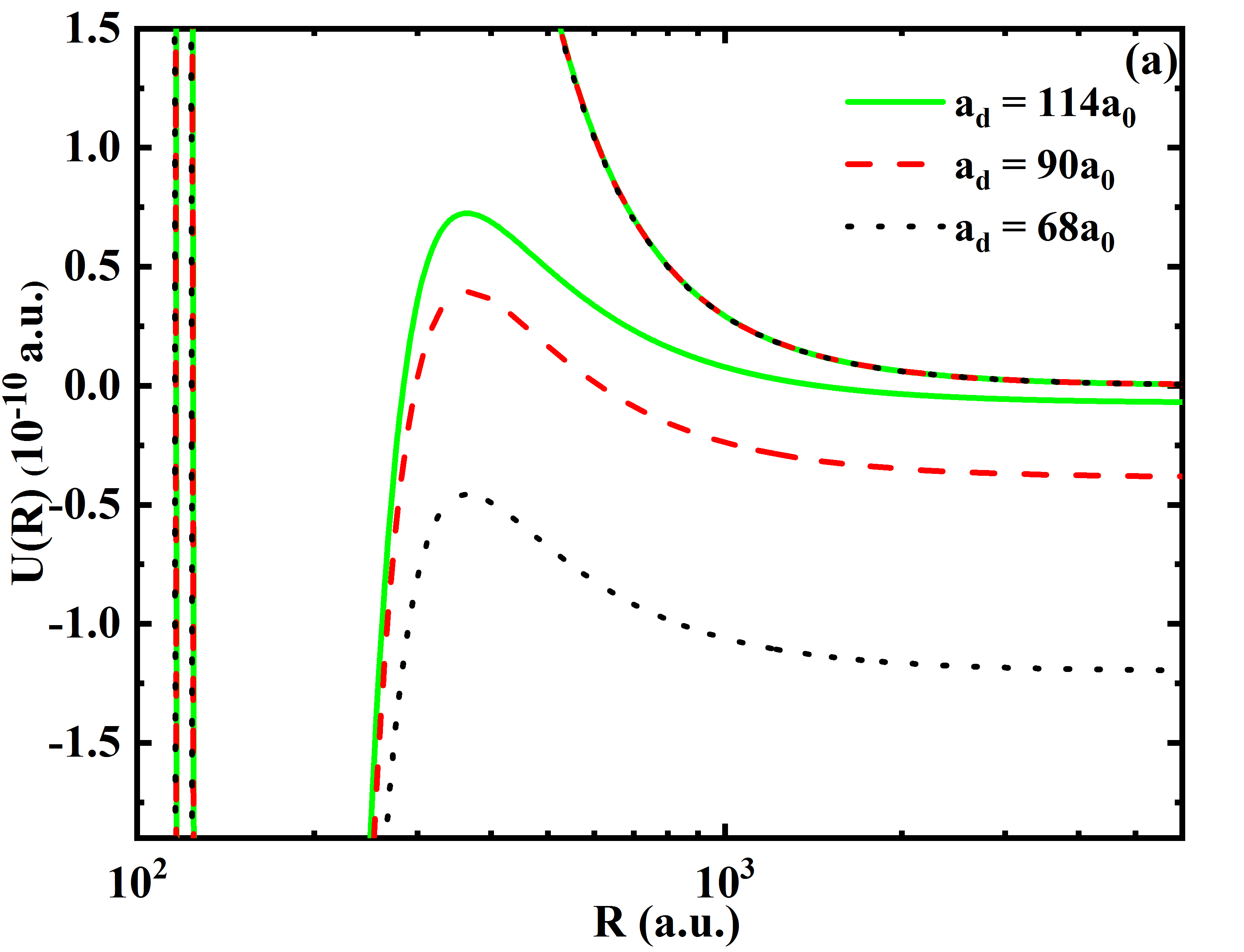}
\label{fig4a}
}
%\quad
\subfigure{
\includegraphics[width=7cm]{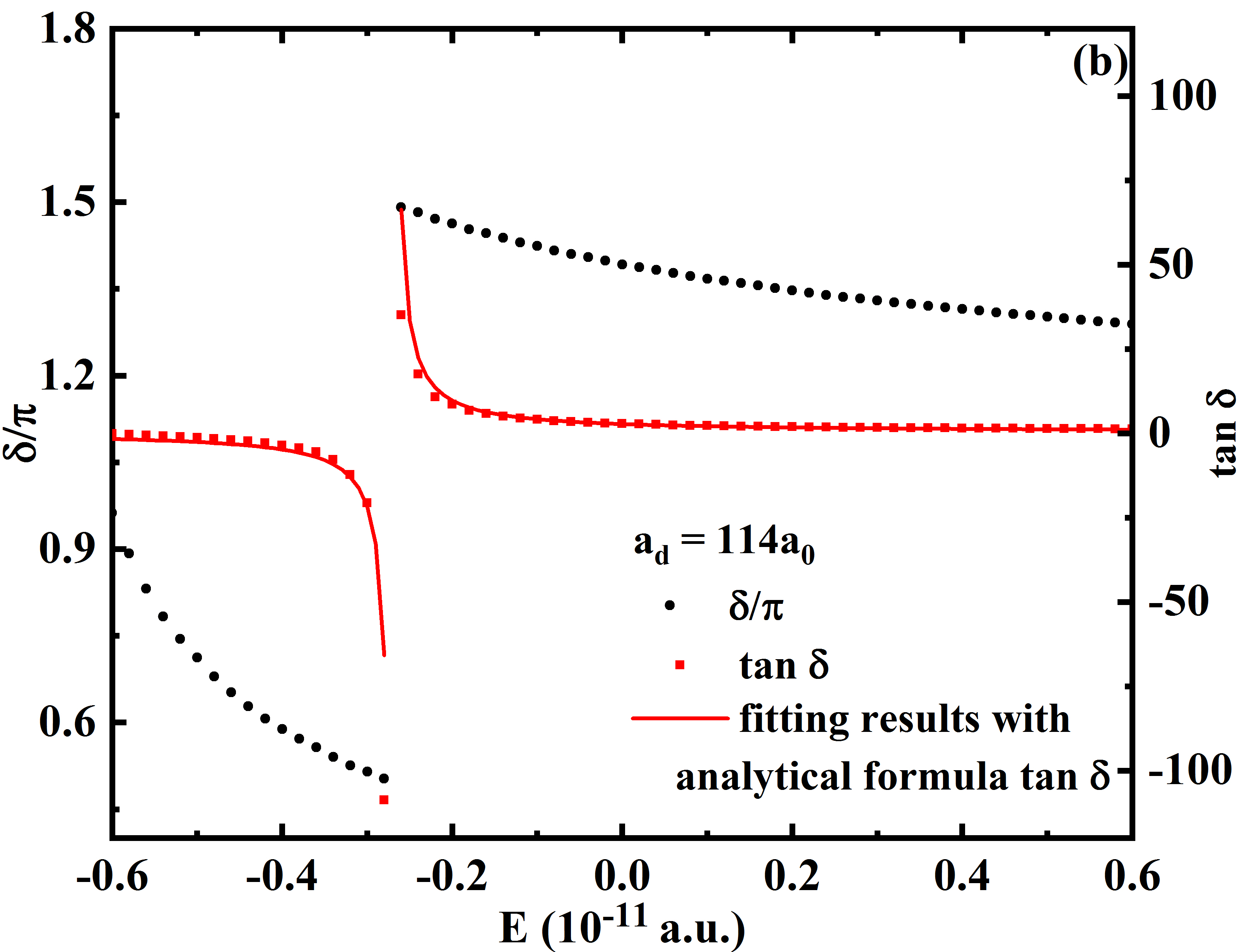}
\label{fig4b}
}
\subfigure{
\includegraphics[width=7cm]{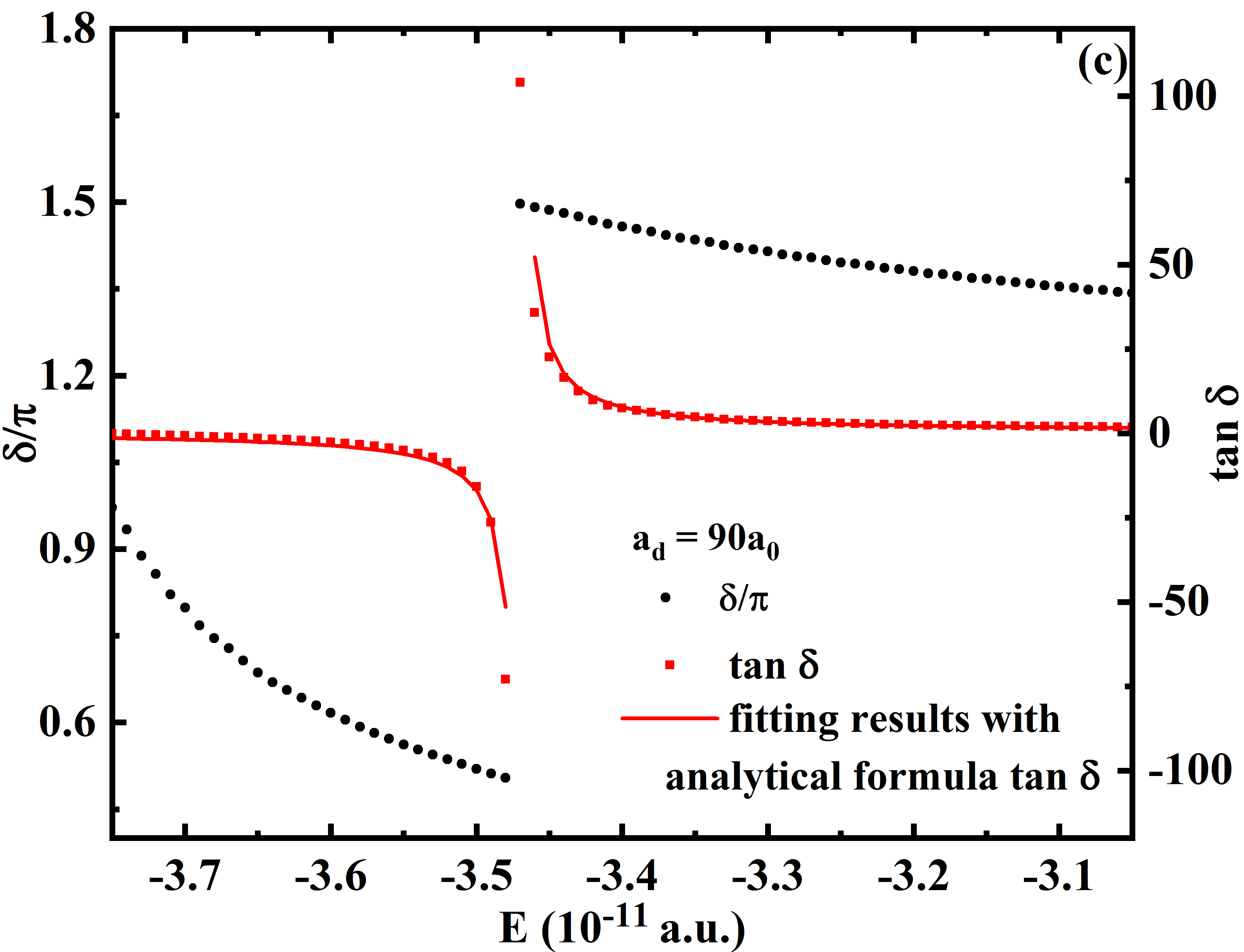}
%\caption{fig1}
\label{fig4c}
}
%\quad
\subfigure{
\includegraphics[width=7cm]{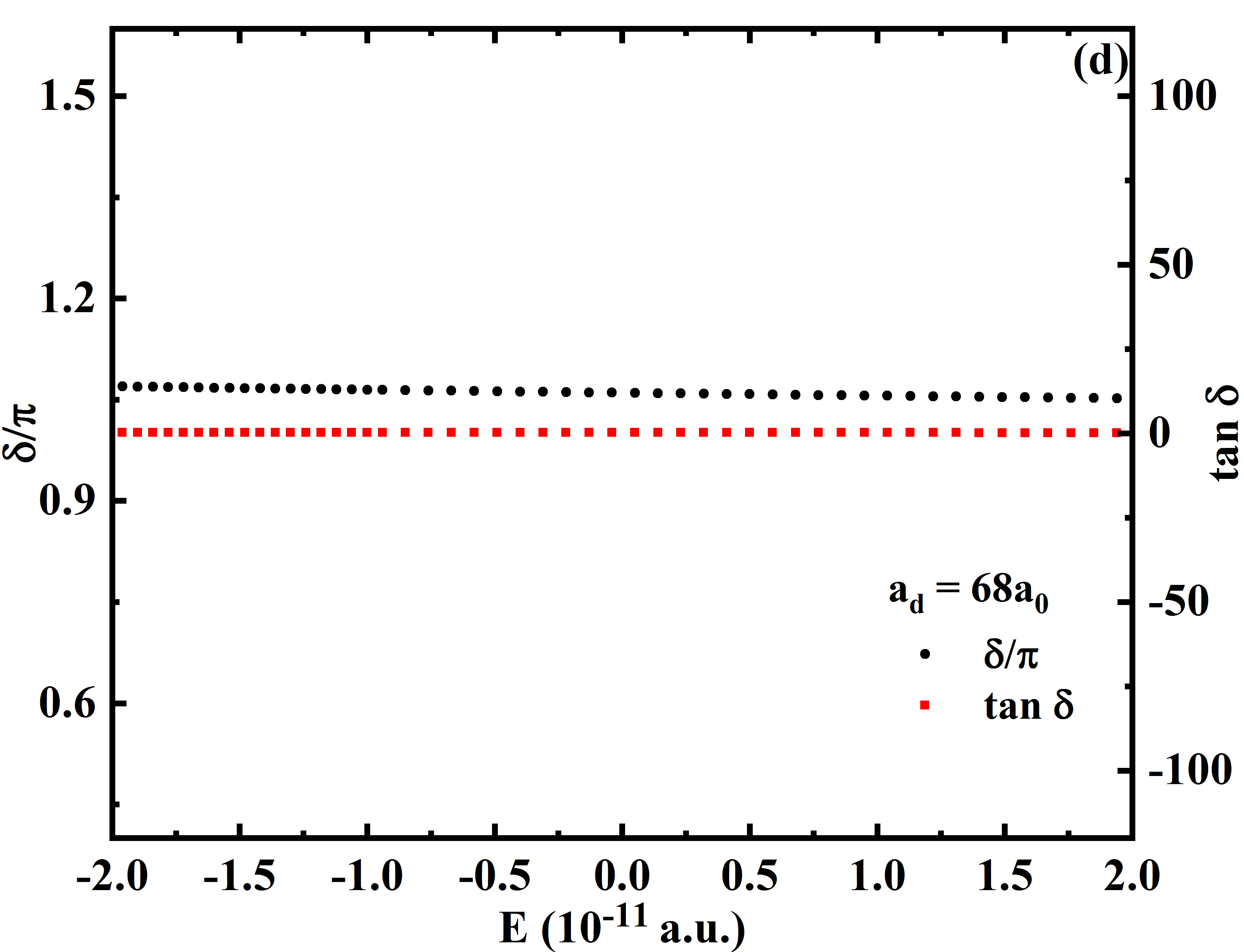}
\label{fig4d}
}
\caption{(Color online)(a) Lowest entrance and highest recombination channels for the three positive $^{85}$Rb\,-$^{87}$Rb d-wave scattering lengths $a_d$. The parameter of the $^{87}$Rb\,-$^{87}$Rb interaction is adjusted to point II in Fig.\;\ref{fig1}. (b), (c) and (d) are the corresponding atom-dimer scattering phase shifts $\delta$ and $\tan\delta$ and the analytical formula for $\tan\delta$ associated with the highest recombination channel as a function of energy. The red sold lines in (b) and (c) are the fitting results by the analytical formula $\tan\delta=\frac{\tan\delta_{bg}-\frac{\Gamma}{2(E-E_R)}}{1+\tan\delta_{bg}\frac{\Gamma}{2(E-E_R)}}$.}
\label{fig4}
\end{figure}

For the positive $^{85}$Rb\,-$^{87}$Rb d-wave scattering length case, the important feature is the broad avoided crossing between the lowest entrance channel and the highest recombination channel, as shown in Fig.\;\ref{fig2c}. The shallowest recombination channel has an attractive well followed by a repulsive barrier at a larger distance. If the barrier is high enough, the atom and dimer could be trapped inside. Such almost bound states above the threshold of the potential are the shape resonance states. Figure\;\ref{fig4a} shows the lowest entrance channel and the shallowest recombination channel of the $^{85}$Rb\,-$^{87}$Rb\,-$^{87}$Rb system for different $^{85}$Rb\,-$^{87}$Rb d-wave scattering lengths with the parameter $\gamma_{ij}$ of the $^{87}$Rb\,-$^{87}$Rb interaction adjusted to point II in Fig.\;\ref{fig1}. The two-body threshold moves towards the three-body breakup threshold with an increase in the $^{85}$Rb\,-$^{87}$Rb d-wave scattering length. Thus, when the top of the barrier is above the collision energy, recombination will be suppressed by this extra barrier, except possibly at energies that match atom-dimer three-body shape resonances behind this barrier. The existence of three-body shape resonance will lead to more or fewer sudden jumps of the phase shift by $\pi$. The phase shift is well described by the analytical expression:
\begin{equation}
\delta_{l}(E)=\delta_{\mathrm{bg}}-\arctan \left(\frac{\Gamma / 2}{E-E_{\mathrm{R}}}\right),
\end{equation}
where $E_R$ is the resonance position, $\Gamma$ is the resonance width, and $\delta_{bg}$ is a smoothly energy-dependent background phase shift\;\cite{PhysRevA.19.920}.
The phase shift $\delta_f$ for atom-dimer elastic scattering is related to the diagonal S-matrix element by the formula
\begin{equation}
S_{f\leftarrow f}^{0+}=\exp(2i\delta_f).
\end{equation}
The subscript f distinguishes the recombination channels.
In Figs.\;\ref{fig4b}-\ref{fig4d}, we plot the atom-dimer scattering phase shift $\delta$, $\tan \delta$ and and the analytical formula for $\tan\delta$ associated with the highest recombination channel as a function of energy. In Figs.\;\ref{fig4b} and \ref{fig4c}, the resonances appear as a jump of the atom-dimer scattering phase shift $0.993\pi$, which implies that shape resonances exist when the $^{85}$Rb\,-$^{87}$Rb scattering length is $a_d=114a_0$ and $a_d=90a_0$. By fitting the $\tan \delta$ points with the analytical expression, we obtain resonance positions of $-2.7\times10^{-12}$ (Fig.\;\ref{fig4b}) and $-3.5\times10^{-11}$ (Fig.\;\ref{fig4c}).

\subsection{Three-body recombination rates}
\label{subsectionB}
\begin{figure}[htbp]
\centering
\subfigure{
\includegraphics[width=7cm]{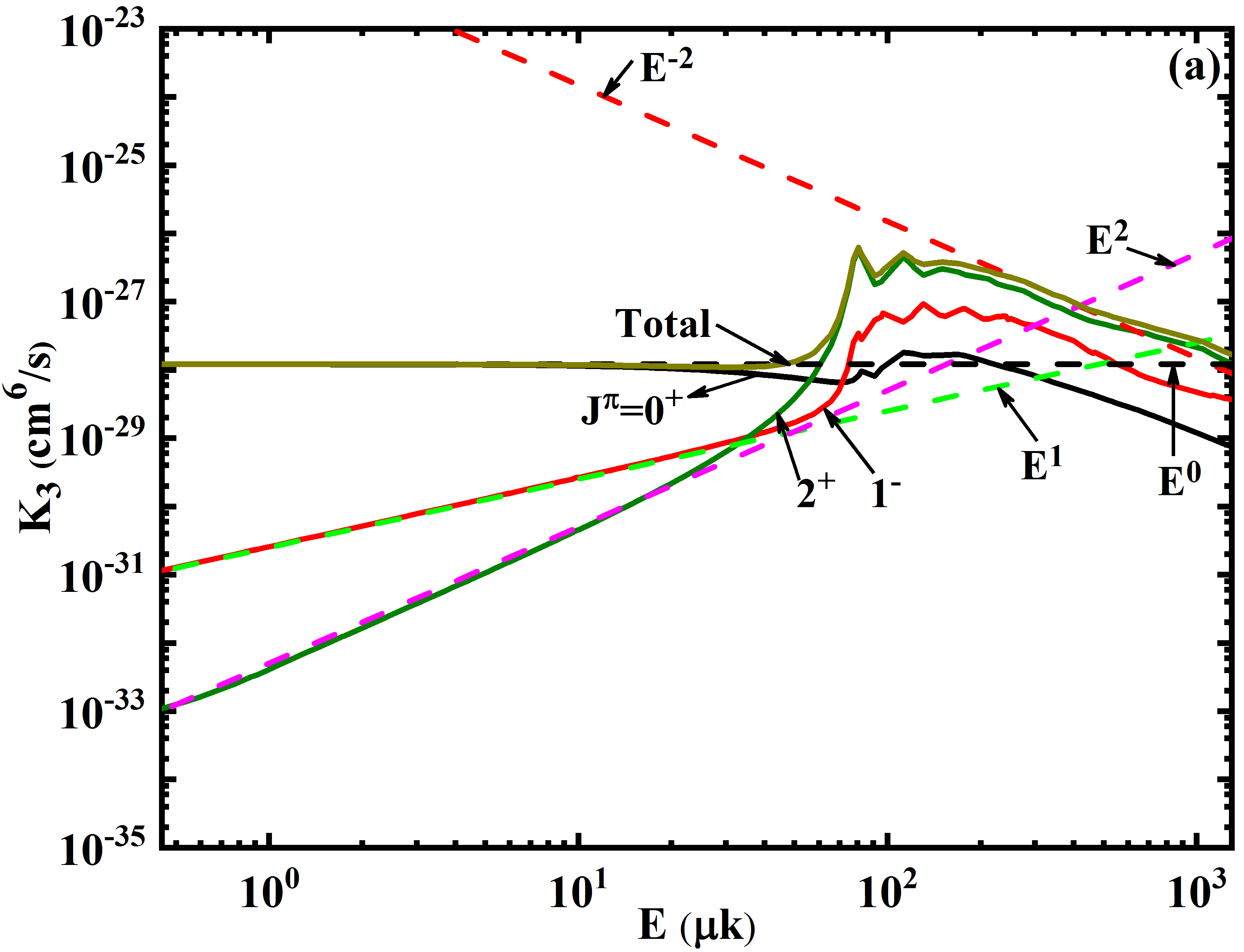}
%\caption{fig1}
\label{fig5a}
}
%\quad
\subfigure{
\includegraphics[width=7cm]{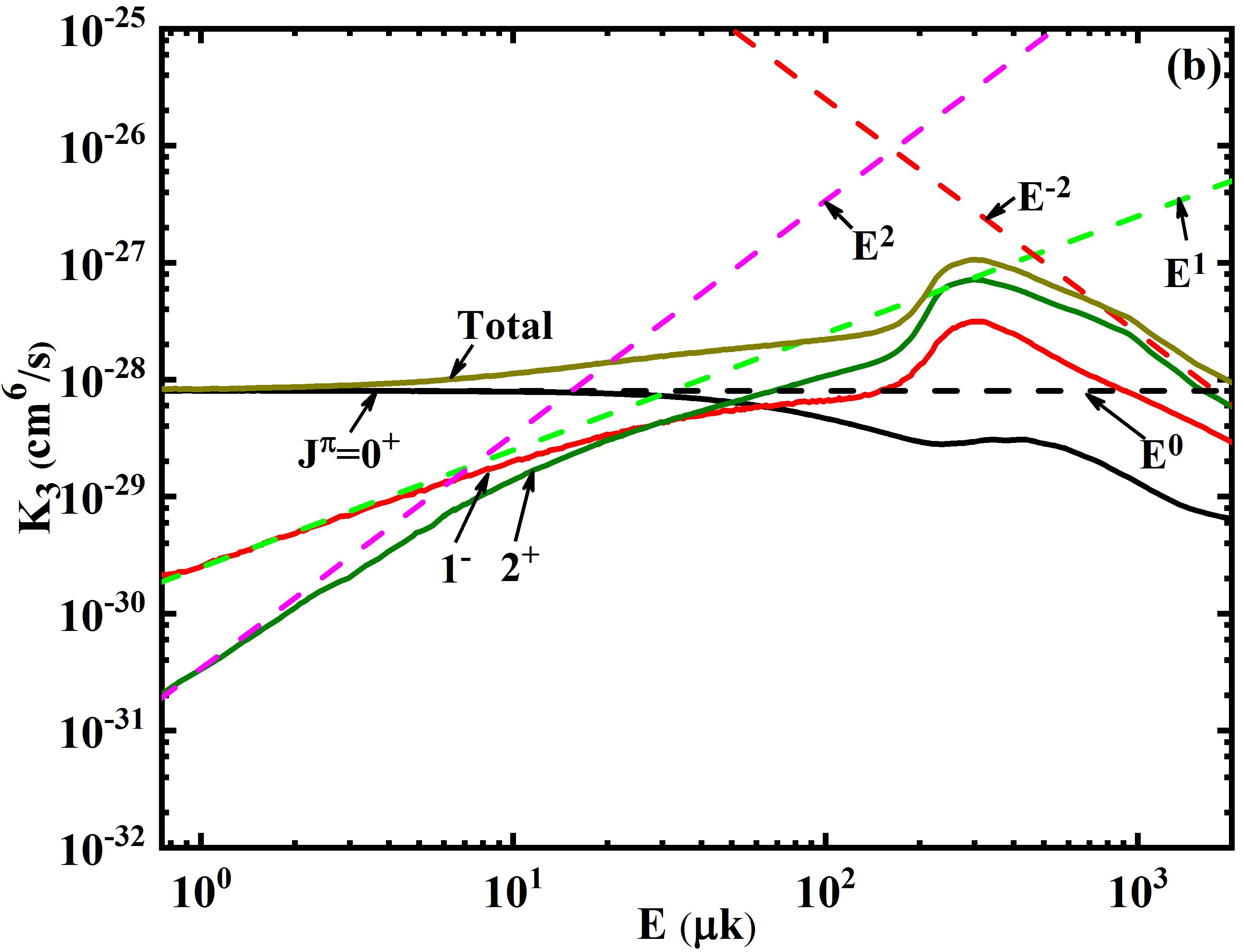}
\label{fig5b}
}
\caption{(Color online) Partial rates $K_3^{J\Pi}$ and their total $K_3$ for $J^\Pi=0^+, 1^-$ and $2^+$ symmetries as a function of the collision energy when the $^{85}$Rb\,-$^{87}$Rb d-wave scattering length is fixed at $a_d=-115a_0$ (Fig. (a)) and $a_d=90a_0$ (Fig. (b)) with the parameter of the $^{87}$Rb\,-$^{87}$Rb interaction adjusted to point II in Fig.\;\ref{fig1}. The dashed lines represent the threshold laws or unitary limit. }
\label{fig5}
\end{figure}

The partial rates $K_3^{J\Pi}$ and total $K_3$ for $J^\Pi=0^+, 1^-$ and $2^+$ symmetries as a function of the collision energy with the $^{85}$Rb\,-$^{87}$Rb d-wave scattering length fixed at $a_d=-115a_0$ (Fig.\;\ref{fig5a}) and $a_d = 90 a_0$ (Fig.\;\ref{fig5b}) are shown in Fig.\;\ref{fig5}. The parameter of the $^{87}$Rb\,-$^{87}$Rb interaction is adjusted to point II in Fig.\;\ref{fig1}. In the zero-energy limit, the recombination rate obeys the threshold behavior $K^{J\Pi}_3 \sim E^{\lambda_{min}}$, where $\lambda_{min}$ is the minimum value of $\lambda$ in Eq.(\ref{43}). For $J^{\Pi}= 0^{+},1^{-},2^{+}$, we have $\lambda_{min} = 0, 1, 2$ in the $^{85}$Rb\,-$^{87}$Rb\,-$^{87}$Rb system\;\cite{PhysRevA.65.010705,PhysRevA.78.062701}, that is, the $J^\Pi=0^+$ partial rate increases, like $E^0$, from the threshold, while the rates $1^-$ and $2^+$ behave as $E^1$ and $E^2$, respectively. At high collision energies, $K_3$ decreases as $E^{-2}$, required by unitary. For the positive d-wave scattering length, Fig.\;\ref{fig5b} shows that the Wigner threshold law holds only at small energies $E< 10 ~\mu K$. Note that the Wigner threshold regime can be characterized as energies smaller than the smallest energy scale, which is typically a molecular binding energy\;\cite{PhysRevA.80.062702}. In this case, when $a_d=90a_0$, the newly formed d-wave dimer binding energy is approximately $12~\mu K$, which shows rough agreement with the attained threshold regime of $E< 10 ~\mu K$.

Although the $J^\Pi=0^+$ case is predicted to be the dominant symmetry in the zero collision energy limit, other symmetries may contribute substantially. Therefore, studying the energy-dependent partial recombination rates that correspond to various symmetries is interesting. Figure\;\ref{fig5} shows that the contributions from $J^\Pi=1^-$ and $J^\Pi=2^+$ partial waves are significant when the energy exceeds $40~\mu K$.

\subsection{Nonnegligible role of the intraspecies interaction in the heteronuclear system}
\label{subsectionC}
In a real ultracold atomic system, the inter- and intraspecies interactions are generally not controlled independently, and thus, a finite intraspecies scattering length exists. For the $^{85}$Rb\,-$^{87}$Rb\,-$^{87}$Rb system, near the interspecies d-wave Feshbach resonance, $^{87}$Rb\,-$^{87}$Rb interact with each other through a smaller s-wave scattering length $a_s=100a_0$. As shown in Fig\;\ref{fig1}, $^{87}$Rb\,-$^{87}$Rb has an s-wave bound state at point II but no s-wave bound state at point III with the same $^{87}$Rb\,-$^{87}$Rb s-wave scattering length.

\begin{figure}[htbp]
\centering
\label{fig6}
%\vskip-5.5ex
\renewcommand{\thesubfigure}{(a\arabic{subfigure})}
\setcounter{subfigure}{0}
\subfigure{
\includegraphics[width=5.2cm]{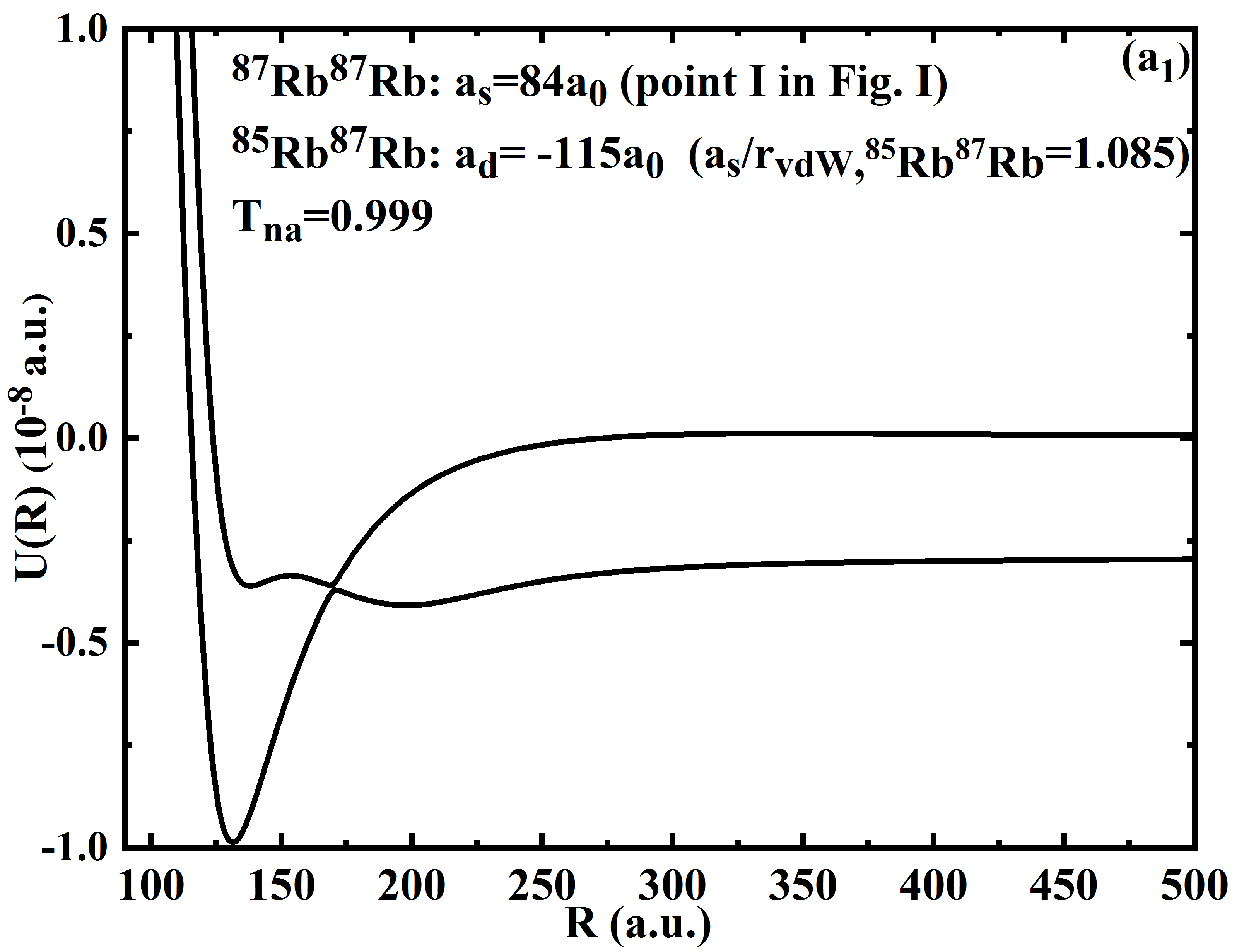}
\label{fig6a1}
}
\subfigure{
\includegraphics[width=5.2cm]{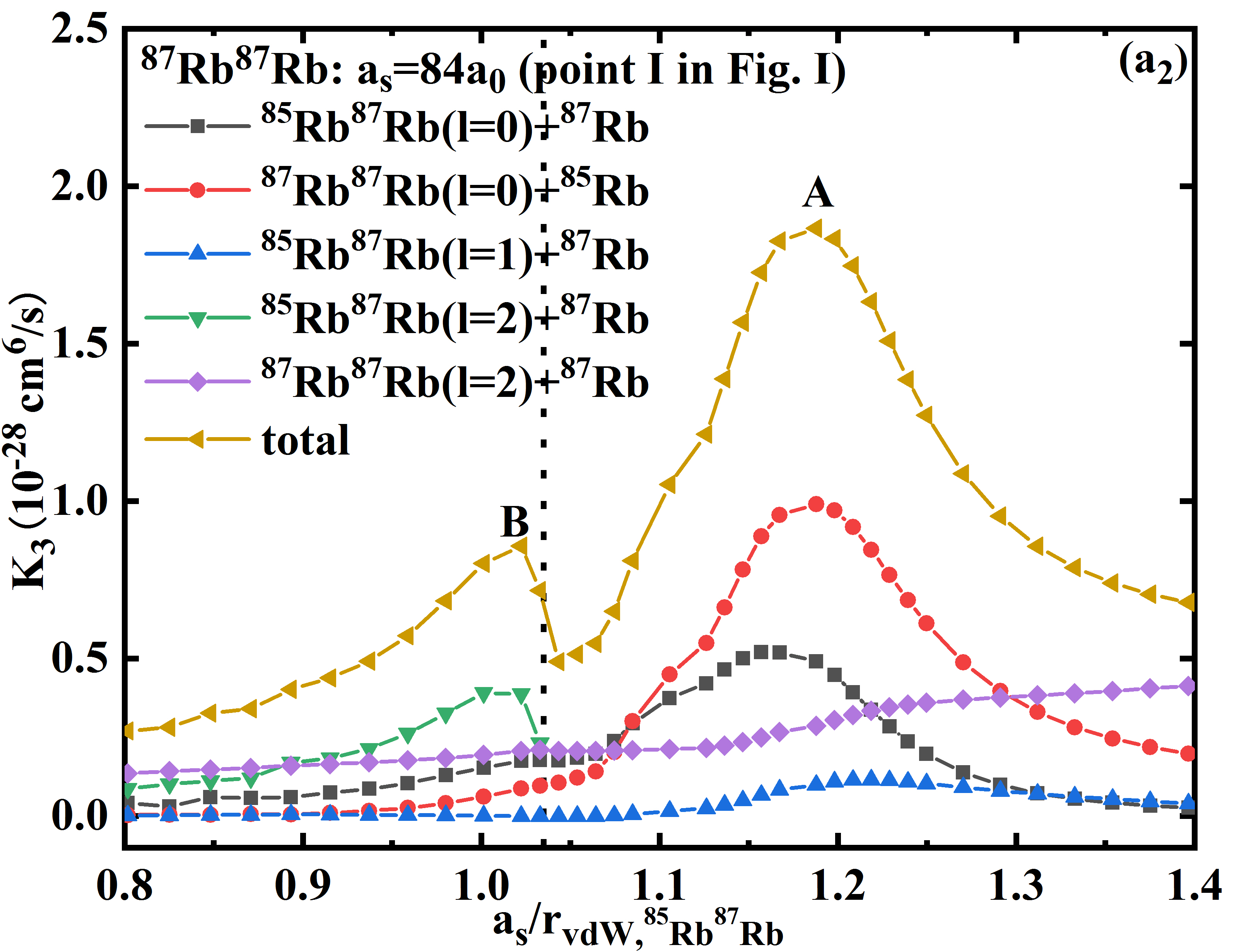}
\label{fig6a2}
}
%\quad
\renewcommand{\thesubfigure}{(b\arabic{subfigure})}
\setcounter{subfigure}{0}
\subfigure{
\includegraphics[width=5.2cm]{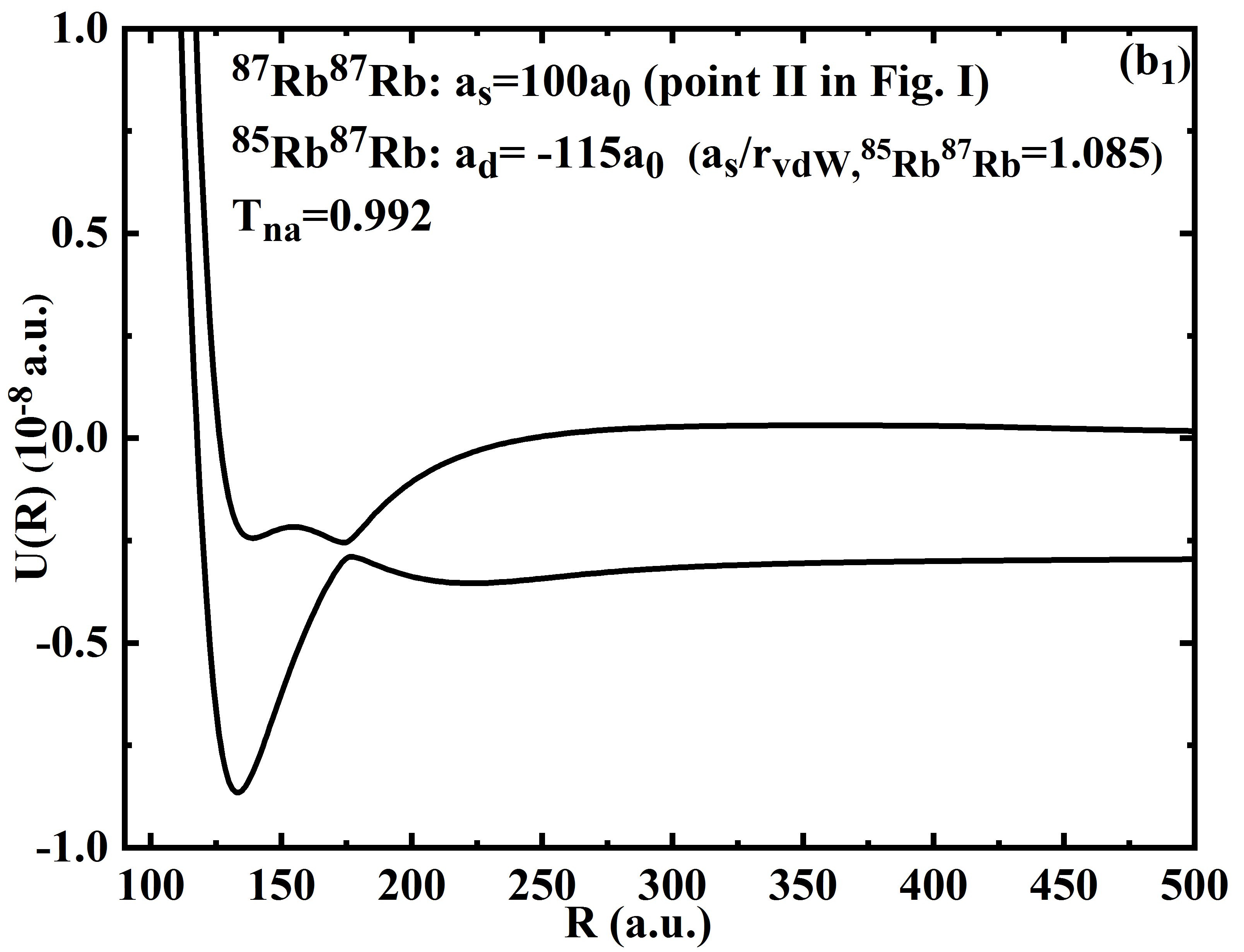}
\label{fig6b1}
}
%\quad
\subfigure{
\includegraphics[width=5.2cm]{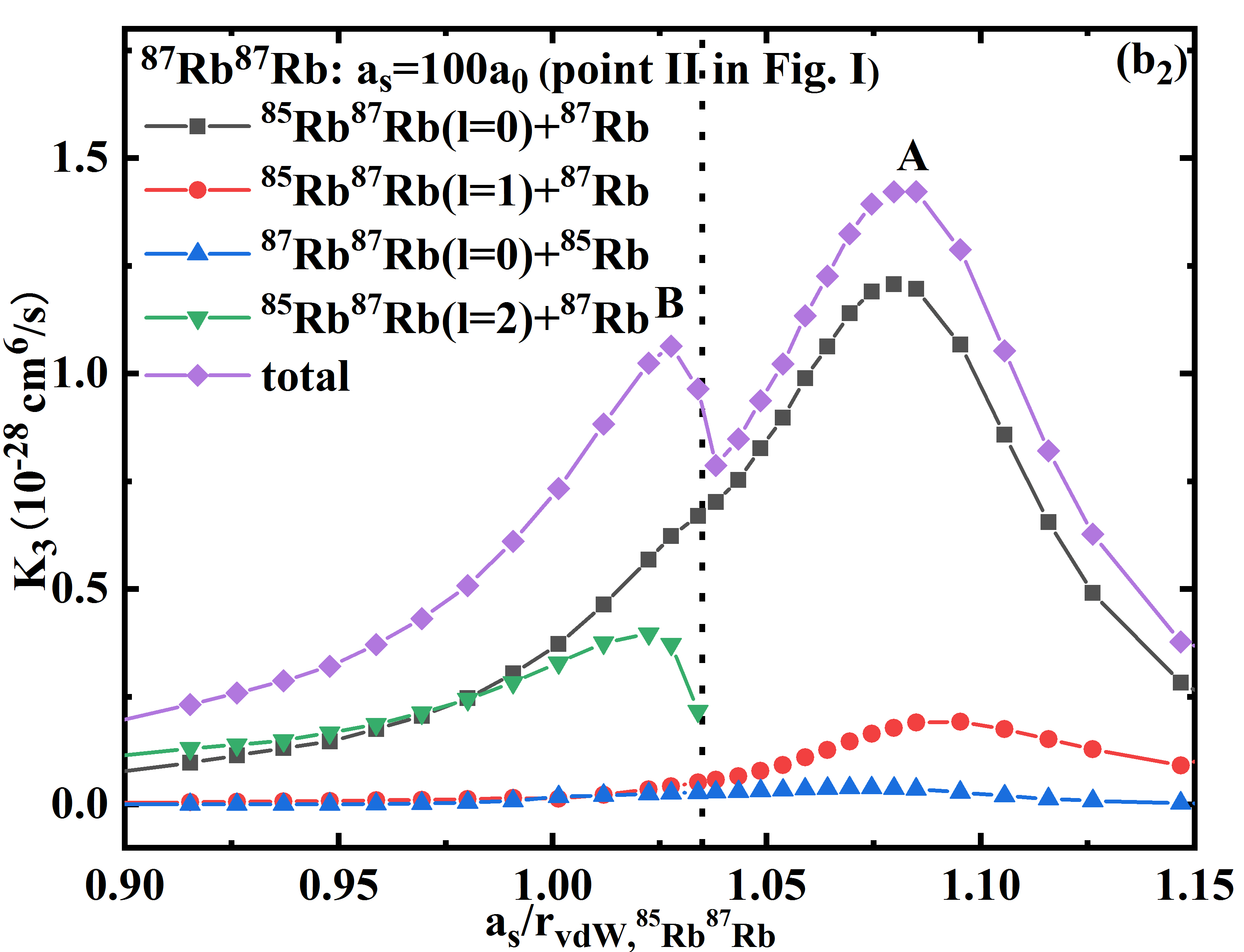}
\label{fig6b2}
}
\renewcommand{\thesubfigure}{(c\arabic{subfigure})}
\setcounter{subfigure}{0}
\subfigure{
\includegraphics[width=5.2cm]{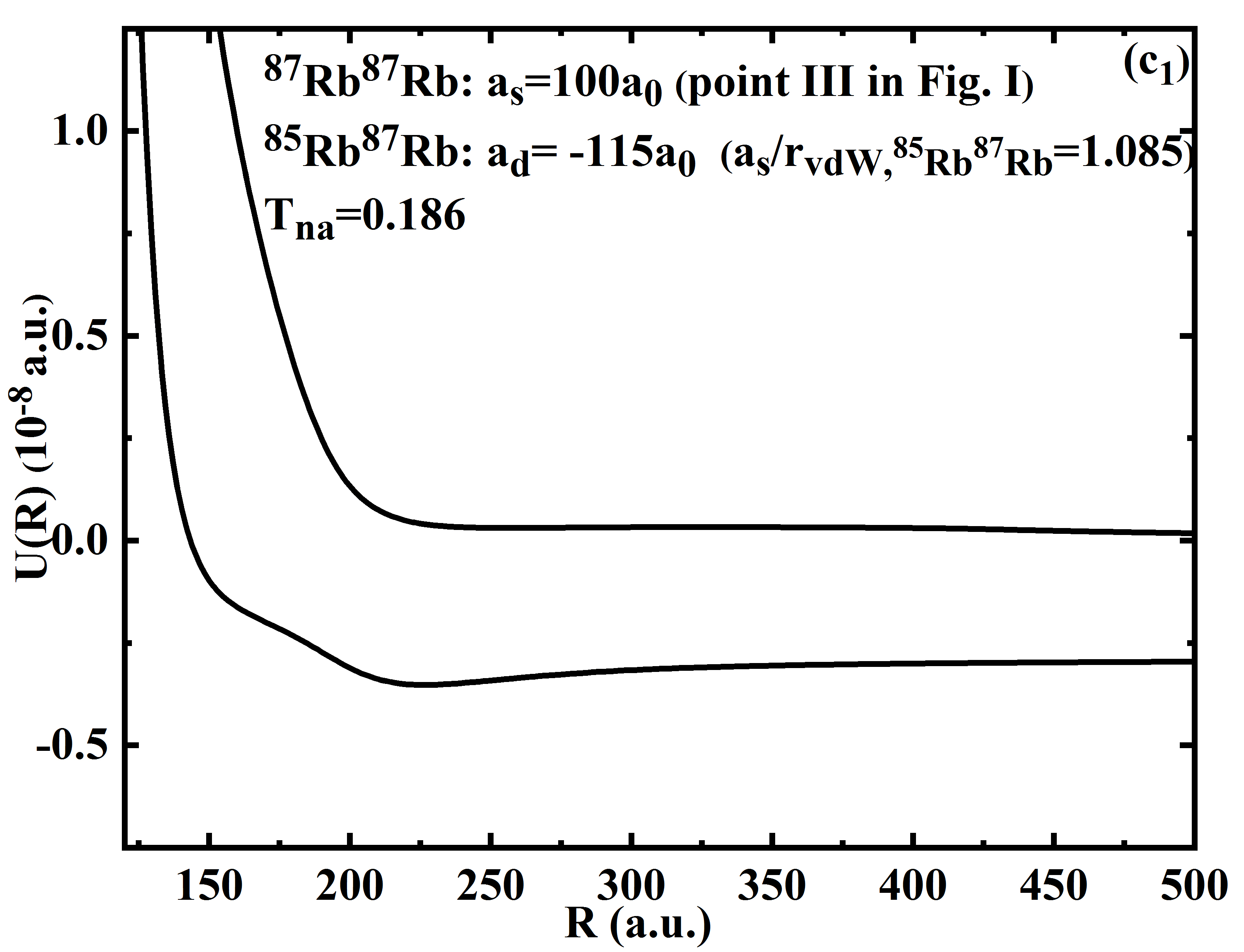}
\label{fig6c1}
}
%\quad
\subfigure{
\includegraphics[width=5.2cm]{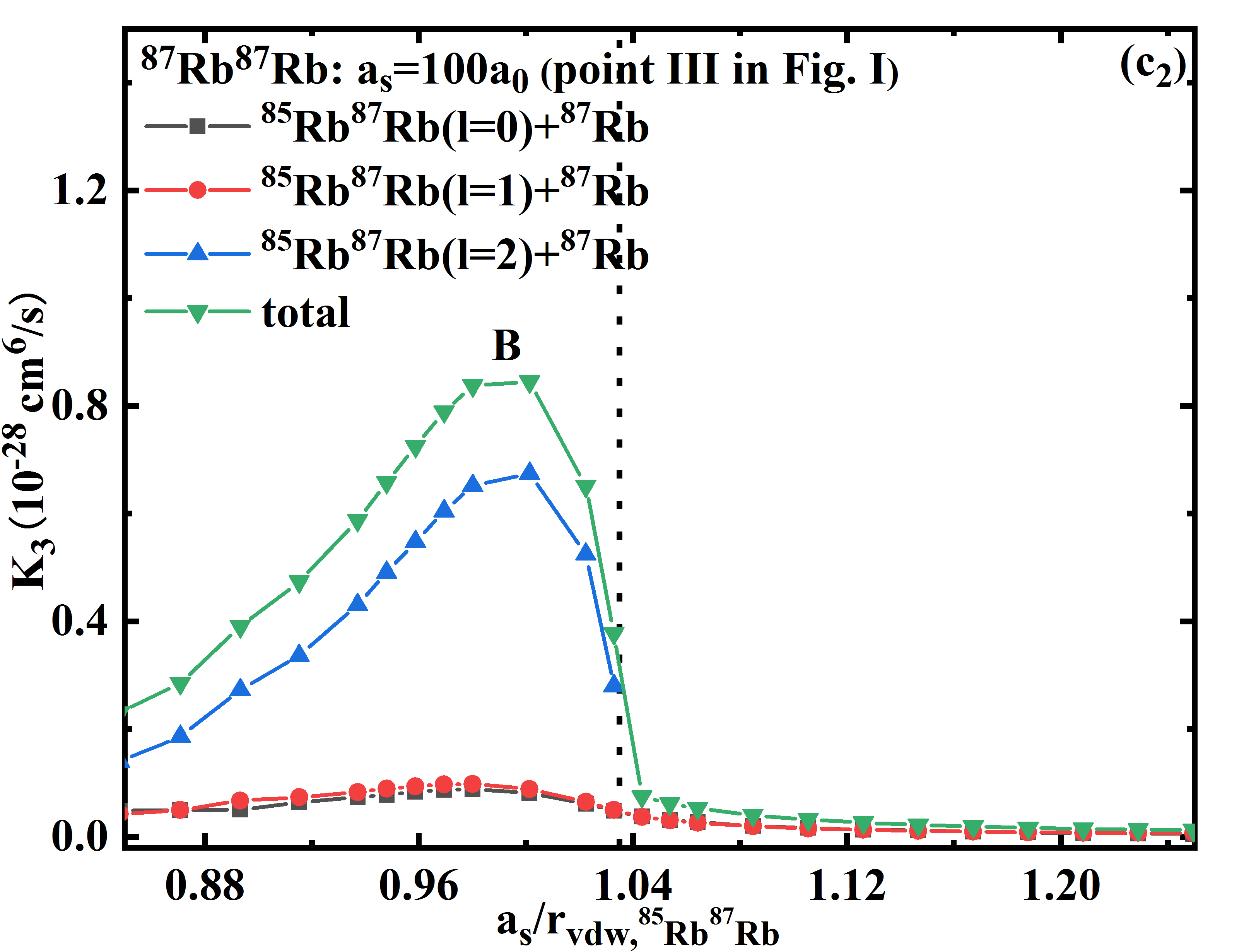}
\label{fig6c2}
}
%\quad
\renewcommand{\thesubfigure}{(d\arabic{subfigure})}
\setcounter{subfigure}{0}
\subfigure{
\includegraphics[width=5.2cm]{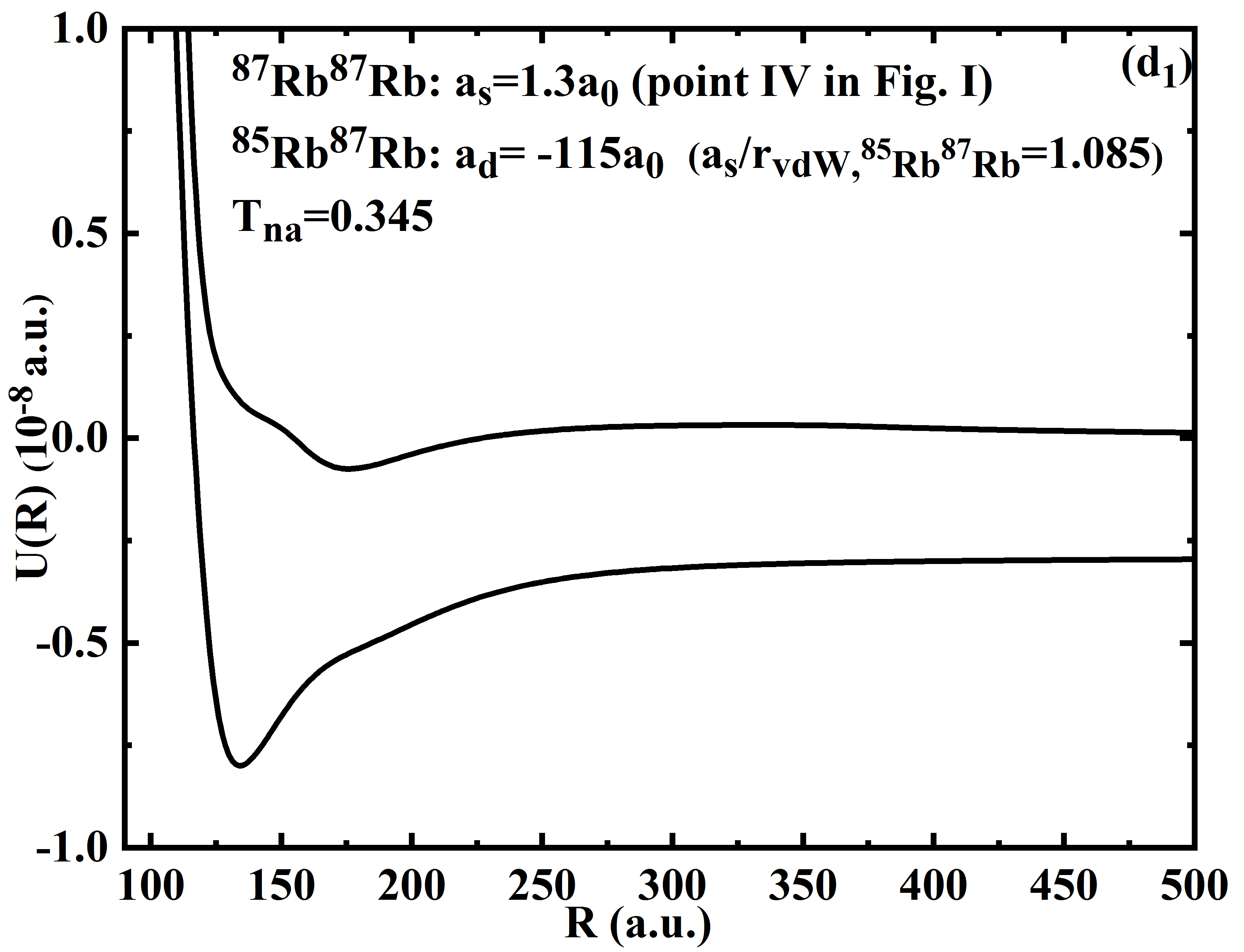}
\label{fig6d1}
}
%\quad
\subfigure{
\includegraphics[width=5.2cm]{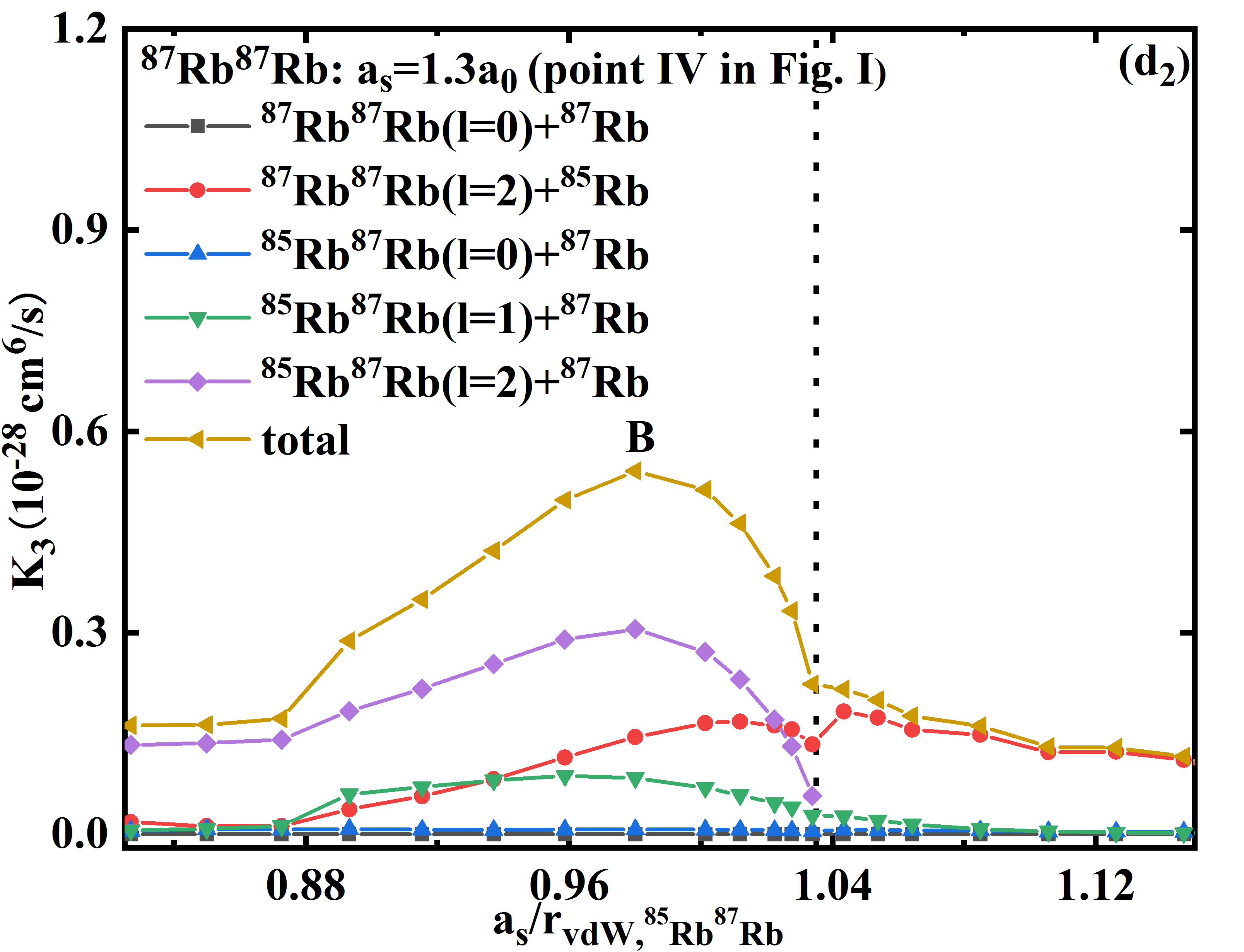}
\label{fig6d2}
}
\caption{(Color online) The first entrance channel and highest recombination channel hyperpotential curves and corresponding Landau-Zener parameter $T_{na}$ for the same $^{85}$Rb\,-$^{87}$Rb d-wave scattering length $a_d=-115a_0$ are shown in (a1), (b1), (c1) and (d1). The TBR rate $K_3$ varies with the $^{85}$Rb\,-$^{87}$Rb interaction when the $^{87}$Rb\,-$^{87}$Rb interaction is fixed at different points in Fig.\;\ref{fig1} are shown in (a2), (b2), (c2) and (d2).The vertical black dotted lines in (a2), (b2), (c2) and (d2) indicate the position of $^{85}$Rb\,-$^{87}$Rb d-wave resonances.}
\end{figure}

To examine how the $^{87}$Rb\,-$^{87}$Rb interaction influences the TBR, we plot the hyperspherical potential curves for the first entrance channel and the highest-lying recombination channel for the same $^{85}$Rb\,-$^{87}$Rb\, d-wave scattering length $a_d = -115\,a_0$ but different $^{87}$Rb\,-$^{87}$Rb\. Interaction details are shown in Figs.\;\ref{fig6a1}, \ref{fig6b1}, \ref{fig6c1} and \ref{fig6d1}. The interaction details of $^{87}$Rb\,-$^{87}$Rb greatly affect the coupling between the lowest entrance channel and the highest recombination channel. The Landau-Zener parameter $T_{ij}$, which estimates the nonadiabatic transition probabilities\;\cite{CLARK1979295}, can quantitatively reflect this coupling strength and be calculated by
$T_{ij}=e^{-\delta_{ij}}=e^{-\frac{\pi \Delta^2_{ij}}{2\alpha_{ij}\nu}}$,
where $\Delta_{ij}=U_i-U_j$ is evaluated in the transition region and $\alpha_{ij}$ is obtained from P-matrix analysis.

When $^{87}$Rb\,-$^{87}$Rb are in d-wave resonance ($^{87}$Rb\,-$^{87}$Rb s-wave scattering length $a_s=84\,a_0$), point I in Fig.\;\ref{fig1}, the Landau-Zener parameter approaches $1$, as shown in Fig.\;\ref{fig6a1}, implying that a nonadiabatic transition occurs and the potential well of the recombination channel deepens. A d-wave-related trimer state (A peak in Fig.\;\ref{fig6a2}) can thus be supported, which will enable enhancement of the TBR rate. When we plot the total and partial $J^{\Pi} = 0^{+}$ TBR rates as a function of the $^{85}$Rb\,-$^{87}$Rb d-wave scattering length at fixed $^{87}$Rb\,-$^{87}$Rb s-wave scattering length $a_s= 84\,a_0$, the total rate exhibits two clear enhancements, labeled ``A`` and ``B``, as shown in Fig.\;\ref{fig6a2}. This phenomenon was predicted by Wang \textit{et al.}\;\cite{PhysRevA.86.062511} in a three-identical-boson system. Peak A corresponds to the d-wave-related trimer state across the collision threshold. According to our analyses of the influence of the $^{87}$Rb\,-$^{87}$Rb interaction on the coupling between the first entrance channel and the highest-lying recombination channel, enhancement A will shift or disappear depending on the interaction details of the two identical atoms in the heteronuclear system. This result can be demonstrated in the following results of $K_3$, when $^{87}$Rb\,-$^{87}$Rb is not exactly in d-wave resonance.

Figures\;\ref{fig6b1} and \ref{fig6c1} show the hyperspherical potential curves with the same $^{85}$Rb\,-$^{87}$Rb two-body scattering length $a_d=-115 a_0$, as shown in Fig.\;\ref{fig6a1}, while $^{87}$Rb\,-$^{87}$Rb interact through s-wave scattering length $a_s= 100 a_0$ with different interaction details. We focus on the case in which the parameter of the $^{87}$Rb\,-$^{87}$Rb interaction is adjusted to point II in Fig.\;\ref{fig1}, where $^{87}$Rb\,-$^{87}$Rb has one s-wave bound state
near the d-wave resonance. In this case, we find that the position of enhancement A shifts from $1.19\,r_{\scriptscriptstyle\textsl{vdW}} $ to $1.08\,r_{\scriptscriptstyle\textsl{vdW}} $ ($r_{\scriptscriptstyle\textsl{vdW}} $ is the van der Waals length between $^{85}$Rb and $^{87}$Rb) compared to the case in which the homonuclear atoms are in d-wave resonance, as shown in Fig.\;\ref{fig6a2}. When the parameter $\gamma_{ij}$ is changed to point III in Fig.\;\ref{fig1}, where the interaction of $^{87}$Rb\,-$^{87}$Rb is away from the d-wave resonance, the coupling between the first entrance channel and the highest-lying recombination channel weakens. Figure\;\ref{fig6c1} shows that the Landau-Zener parameter is $0.186$ in this case, implying that an adiabatic transition occurs. As a result, the potential well in the recombination channel may not be deep enough to support the trimer state, leading to the absence of enhancement A in Fig.\;\ref{fig6c2}. Fig.\;\ref{fig6d1} show the hyperspherical potential curves, where the parameter $\gamma_{ij}$ is changed to point IV in Fig.\;\ref{fig1}, where $^{87}$Rb\,-$^{87}$Rb interact through s-wave scattering length $a_s= 1.3 a_0$ away from the d-wave resonance. We note that the Landau-Zener parameter is $0.345$ and enhancement A also disappears in this case.

Peak B formed after the d-wave dimer became bound. Our analysis in\;\ref{subsectionA} shows that this enhancement corresponds to the three-body shape resonance. Its position is also affected by the interaction details of the two homonuclear atoms, as shown in Figs.\;\ref{fig6a2}, \ref{fig6b2} , \ref{fig6c2} and \ref{fig6d2}.
Our results demonstrate that the intraspecies interaction has a significant role in determining the TBR in the heteronuclear system.

\section{Conclusions}
In summary, we have investigated the TBR rate for the heternuclear atomic system near the interspecies d-wave Feshbach resonance. The $^{85}$Rb\,-$^{87}$Rb-$^{87}$Rb system is chosen as an example and calculations are based on the Lennard-Jones model potential for the Rb-Rb interaction. The TBR rates are obtained using quantum calculations in the frame of the hyperspherical coordinates, which are based on a combination of the SVD method, traditional hyperspherical adiabatic method and
R-matrix propagation method. Our study reveals two different mechanisms of recombination rate enhancement for positive and negative $^{85}$Rb\,-$^{87}$Rb d-wave scattering lengths. When the $^{85}$Rb\,-$^{87}$Rb d-wave scattering length is positive and large, a loosely bound dimer is produced, and enhancement occurs due to the existence of three-body shape resonance. We have identified two such shape resonances on the positive $^{85}$Rb\,-$^{87}$Rb d-wave scattering length side. For the case in which $^{87}$Rb\,-$^{85}$Rb interact via a negative d-wave scattering length, the coupling between the lowest entrance channel and the highest recombination channel is crucial to the formation of the three-body state. When the coupling strengthens, a nonadiabatic transition occurs, which deepens the potential well in the highest recombination channel and thus supports the three-body state. The enhancement on the negative interspecies scattering length side corresponds to the three-body state crossing the three-body threshold.

In addition, we investigated the influence of the finite $^{87}$Rb\,-$^{87}$Rb interaction on the recombination enhancement. With the same $^{87}$Rb\,-$^{87}$Rb s-wave scattering length $a_s=100 a_0$, when the interaction is near the d-wave resonance, the coupling between the lowest entrance channel and the highest recombination channel strengthens, enhancing the recombination rate. However, if the parameter of the $^{87}$Rb\,-$^{87}$Rb interaction is adjusted to a point away from the d-wave resonance, then the enhancement will disappear. Moreover, our study reveals that the intraspecies interaction affects the $^{85}$Rb\,-$^{87}$Rb d-wave scattering length values at which the enhancement appears. Our results have confirmed the main results of Ref.\;\cite{PhysRevA.86.062511} for the homonuclear case and provide numerical evidence that the TBR rates in heteronuclear systems are more complex than those in homonuclear systems.

\section{Acknowledgments}
We thank C. H. Greene, Jia Wang, Li You, Meng Khoon Tey and Cui Yue for helpful discussions.
Hui-Li Han was
supported by the National Natural Science Foundation of China under Grants No. 11874391 and No. 11634013 and the National Key Research and Development
Program of China under Grant No. 2016YFA0301503.
Ting-Yun Shi was supported by the Strategic Priority Research Program of the Chinese Academy of Sciences under Grant No. XDB21030300.

%\bibliography{8587}

\end{document}